\documentclass[manuscript]{aastex}

\newcommand{\ctht}{C$_2$H$_2$}

\shorttitle{Gas and Dust in Protoplanetary Disks}
\shortauthors{Pascucci et al.}

\begin{document}


\title{The Different Evolution of Gas and Dust in Disks around Sun-like and Cool Stars}


\author{I. Pascucci\altaffilmark{1}, D. Apai\altaffilmark{2}, K. Luhman\altaffilmark{3}
Th. Henning\altaffilmark{4}, J. Bouwman\altaffilmark{4}, M. R. Meyer\altaffilmark{5},
F. Lahuis\altaffilmark{6}, A. Natta\altaffilmark{7}}
\altaffiltext{1}{Department of Physics and Astronomy, Johns Hopkins University, Baltimore, MD 21218}
\altaffiltext{2}{Space Telescope Science Institute, Baltimore, MD 21218}
\altaffiltext{3}{Department of Astronomy and Astrophysics, The Pennsylvania State University, University Park, PA 16802}
\altaffiltext{4}{Max Planck Institute for Astronomy, Heidelberg, Germany}
\altaffiltext{5}{Steward Observatory, The University of Arizona, Tucson, AZ 85721}
\altaffiltext{6}{SRON Netherlands Institute for Space Research, Groningen, The Netherlands}
\altaffiltext{7}{INAF-Osservatorio di Arcetri, 50125 Firenze, Italy}




\begin{abstract}

Planet formation is profoundly impacted by the properties 
of protoplanetary disks and their central star. However, how 
disk properties vary with stellar parameters remain poorly known. 
Here we present the first comprehensive, comparative Spitzer/IRS 
study of the dust and gas properties of disks around young sun-like stars (K1--M5) and cool 
stars/brown dwarfs (M5--M9). The comparison of these two large samples of over 60 sources reveal 
major differences in the evolution of both the dust and gas components.
 
We report the first detection of organic molecules in disks around brown dwarfs.
The detection rate statistics and the line flux ratios of HCN and C$_2$H$_2$ show a 
striking difference between the two samples, demonstrating a significant under-abundance of HCN relative
to C$_2$H$_2$  in the disk surface of cool stars. We propose this to originate from the
large difference in the UV-irradiation around the two types of sources.
The statistical comparison of the 10\,\micron{} silicate emission features also reveals a
difference between the two samples. Cool stars and brown dwarfs show weaker features arising
from more processed silicate grains in the disk atmosphere.

These findings complement previous indications of flatter disk structures and longer disk lifetimes
around cool stars. Our results highlight important differences in the chemical and physical 
evolution of protoplanetary disks as function of stellar mass, temperature, and radiation field
which should be taken into account in planet formation models. We note that the 
different chemistry of pre-planetary materials in the disk may also influence the bulk composition and 
volatile content of the forming planets. In particular, if exogenous HCN has played a key role 
in the synthesis of prebiotic molecules on Earth as proposed, then prebiotic chemistry may 
unfold differently on planets around cool stars.


\end{abstract}





\section{Introduction}

Circumstellar disks are a natural outcome of the star formation process and the birthplace of planetary systems. Their evolution and their lifetime determine what type of planets can form.
Most studies so far have focused on characterizing protoplanetary disks around young sun-like stars (e.g. \citealt{meyer07} for a review).
It is now well established that most sun-like stars clear out their primordial dust disk in 5\,Myr and only a few percent of them still retain it by an age of 10\,Myr (e.g. \citealt{hernandez07}). The dispersal of the gas component is less well characterized but appears to proceed equally fast \citep{pas06,aguilar06}. This rapid clearing of dust and gas around sun-like stars matches the fast growth of planetesimals and gas dispersal in the protosolar nebula as inferred from meteorites, asteroids, and planets \citep{pascuccitachibana}.
But do disks around stars of different masses evolve similarly? Recent studies hint that disks around cool/late-type M stars, the most typical protoplanetary disks, evolve differently (e.g. \citealt{apai08}). There are at least three major differences. 

First, the dispersal of primordial dust disks seems to operate less efficiently for cool stars.
\citet{carpenter06} find that the disk frequency of low-mass stars ($\sim 0.1-1.2$\,M$_\sun$) is statistically higher than that of higher-mass stars in the 5 Myr-old Upper Sco OB association. Infrared excess studies by
\citet{sterzik04}, \citet{scholz07}, and \citet{riaz08}  all suggest that this trend extends to the brown dwarf regime. Second, there is evidence that disks around cool stars have, on average, flatter disk structures than the more flared disks around sun-like stars (\citealt{pas03,apai04,allers06} but see also the flared disks from \citealt{mohanty04,bouy08}). Third, the dust probed through the 10\,\micron{} silicate emission feature appears to be more processed around cool stars than around sun-like stars of similar age, suggesting more rapid grain growth in cool star disks \citep{apai05,ks06}.

However, there has not been a detailed comparison of the dust properties of disks around sun-like and cool stars and no investigation of their gas content. Here we present the first such comparative study. We will show that there are statistically significant differences between cool and sun-like stars in their gas and dust disk  properties.  First, we will demonstrate that cool stars/brown dwarfs have much weaker continuum-subtracted 10\,\micron{} features than higher-mass stars (Sect.~\ref{res:silicates}) implying different dust populations at the observed disk radii (Sect.~\ref{sect:discuss_dust}). Second, we will show that HCN emission is often present in sun-like star spectra, but absent from the cool star spectra. Similarly, the median sun-like star spectrum has an HCN/\ctht{} flux ratio that is an order of magnitude higher than the median cool star spectrum (Sect.~\ref{res:gas}). We will see that this arises from an under-abundance of HCN in the disk atmosphere of cool stars (Sect.~\ref{discuss:gaslines}). 
These findings highlight how stellar mass, and radiation field affect the physical and chemical evolution of protoplanetary disks including the organic compounds available during planet formation.

\section{Targets, Observations, and Data Reduction}\label{obs}
\subsection{Target Selection}
We compare two large samples of disks around sun-like and cool stars (including brown dwarfs) that have 
Spitzer spectra obtained with the same resolution and very similar signal-to-noise.
The Taurus and the Cha~I star-forming regions provide the best samples of protoplanetary disks to carry out our comparative study. These low-mass star forming regions are at about the same distance and do not contain
early B or O type stars. In addition, their median age ($\sim$1 and 2\,Myr) is the same within our ability to discern (see, e.g. \citealt{kenyon08book} and \citealt{luhman08book} for a review on Taurus and Cha~I).

The sun-like star sample is the same as in \citet{pas08}: it contains 23 single and 21 binary stars 
from the Taurus-Auriga star-forming region  with spectral types between K1--M5,
temperatures between 3,000--5,000\,K, and stellar masses between 0.4--2\,M$_\sun$ (see also \citealt{kh95} for other stellar properties). 
Two-thirds of the binary systems have stellar companions between 0.1\arcsec-1\arcsec , with a mean projected
separation of 0.4\arcsec , or 56\,AU at the distance of the Taurus star-forming region.
 The low-resolution IRS spectra for the sun-like star sample were acquired as part of the IRS/GTO program \citep{furlan06} and have been reduced as described in \citet{pas08} and \citet{bouwman08}. \citet{pas08} showed that there is no statistically significant difference between the strength of the 10\,\micron{} emission
features from the single and binary systems in these two samples. Similarly, they showed that the ensemble of spectral energy distributions is indistinguishable between the single and binary system samples. 

Our cool star sample includes objects with spectral types between M5--M9, stellar temperatures between $\sim$2,500-3,000\,K and masses between $\sim$0.04-0.15\,M$_\sun$ (see Table~\ref{T:photometry} and also \citealt{apai05,luhman07}). These objects were selected from the Cha~I star-forming region following two main criteria: i) reliable spectral classification, and ii) mid--infrared excess emission.  We used the optical/infrared surveys  by 
\citet{comeron00}, \citet{lopez04}, \citet{luhman04}, and 
\citet{luhman05} to identify sources close to or below the substellar boundary.   
We then correlated these sources with the ISO/ISOCAM detections at 6.7 and 14.3\,\micron{} from 
\citet{persi00} and the Spitzer/IRAC 8\,\micron{} fluxes from \citet{luhman08}. 
The ISOCAM--selected sample consists of eight sources: Hn~2, CHXR~15, Cha~H$\alpha$~1, 
Cha~H$\alpha$~9, Cha~H$\alpha$~2, ISO~138, Cha~H$\alpha$~6, and ISO~217. 
These sources were observed in March 2005 as part of our Cycle--1 Spitzer proposal and
were selected to have fluxes larger than 10\,mJy in at least one of the two wavelengths to ensure a
reliable detection of excess emission. Another eleven sources were selected later based
on excess emission in the IRAC 5.8 and/or 8\,\micron{} bandpasses \citep{luhman08}, and  were observed as part of our Cycle--3 Spitzer proposal in August
2006 and March 2007. The IRAC--selected sources are: ESO~H$\alpha$~559, ISO~79, CHSM~9484, Cha~23943,
ISO~147, Cha~9086, T37, ISO~165, ISO~252, Hn~13, and Cha~21330. 

\subsection{Cool Star Sample: Observations}\label{sect:selection}
We used the infrared spectrograph (IRS, \citealt{houck04}) on--board the Spitzer
Space Telescope to perform low--resolution ($R\sim$64--128) spectroscopy of the 
10\,\micron{} silicate emission feature of the selected cool stars. 
We employed for all sources the SL1 module covering the
wavelength range between 7.4--14.5\,\micron. For several sources of the Cycle--1 proposal we also
adopted the LL1 module (19.5 -- 38.0\,\micron) aiming to probe solid--state features at long
wavelengths.  However, because most of the selected brown dwarfs turned out to have flat disk
structures \citep{apai05},  the  signal--to--noise ratio of the LL1 observations is only
modest. This means that we can use the LL1 data only to trace the infrared continuum.  In view of
this finding, we preferred to drop the LL1 module for the Cycle--3 observations. We chose instead
the SL2 module (5.2--7.7\,\micron) to better establish the continuum slope around the 10\,\micron{}
silicate emission feature.  The parameters of the observations are summarized in
Table~\ref{T:Observations}. Most of the  sources were centered in the spectrograph slit using the
high-accuracy PCRS mode (1 $\sigma$ positional uncertainty of 0.4\arcsec). The exceptions are Cha~9086,
ISO~217, and Hn~13 where no suitable optical reference star was available. For these sources, we used
the medium accuracy infrared peak-up array integrated onto the  IRS focal
plate providing a 1 $\sigma$ positional uncertainty of 1\arcsec. 
All peak-up operations were successful and no significant flux loss is expected
from any of our sources. Each source was checked individually to confirm that only one
point source was covered by the slit.  We used fixed cluster offset-mode observations and
grouped nearby targets to reduce the telescope overhead. By using the staring mode observation template, we acquired each target in two nod positions along
the spatial direction of the slit, at 1/3 and
2/3 of the slit length. 
		
\subsection{Data Reduction}\label{S:DataRedu}
The sun-like and cool star samples have been reduced in an identical way, with only very minor differences. 
We have already published the spectra of the sun-like star sample in \citet{pas08}. 
Here we adopted the already reduced spectra from our previous paper and will now only 
discuss the reduction of the cool star sample.

The cool star data have been processed through the SSC pipeline version S15.3.0 and S16.1.0. 
Our goals required a data quality higher than that produced by the SSC pipeline.
We start our data reduction from the so-called {\it droopres} products and use the SMART reduction package \citep{higdon04}, in combination with IDL routines developed for the FEPS Spitzer Science Legacy program \citep{meyer06}. 
The data reduction steps are outlined in detail in \citet{bouwman08} and have been successfully 
used in a series of publications \citep[see, e.g.][]{meyer04,apai05,bouwman06,pas08}. 
In brief, we first subtracted the pairs of imaged spectra acquired along the spatial direction of the slit to 
correct for background emission, stray light, and pixels with anomalous dark current. 
We replaced pixels flagged as bad by the SSC pipeline (and additional 3-4 pixels per frame identified by eye)
 by interpolating over neighboring good pixels. 
Spectra were extracted from the background-subtracted pixel-corrected images using a 6-pixel
fixed-width aperture in the spatial direction.

Because all targets have been acquired with a high or moderate peak-up option (pointing accuracy within 0.4-1\arcsec) and the majority are too faint to perform source profile fitting, we opted to fix the position of the aperture for each spectral order to that derived by the FEPS legacy program. 
After extracting the spectra for each order, nod, and cycle, we computed a mean spectrum for each order and as uncertainty we quoted the 1–$\sigma$ standard deviation of the distribution of the flux densities measured at the given wavelength. We converted our output spectra to flux by applying the spectral response function derived by the FEPS Legacy team \citep{bouwman08}, and propagate the calibration error into the quoted uncertainties. 

\section{Results}\label{S:results}
The SL1 observations detected all cool stars and achieved a sub--mJy accuracy.  We confirm the ISOCAM
excess emission detections in all but two sources: Our Spitzer/IRS observations of Hn~2 and CHXR~15 show flux densities fully consistent with photospheric emission and we identify no
hints of the ISOCAM excess emission reported by \citet{lopez04}.  
It seems likely that these two sources have been missclassified: photospheric emission from sources earlier than our spectral type limit of M6 can fully explain the ISOCAM observations (\citealt{luhman08}, see also Table~\ref{T:photometry}). We excluded these two sources from the following analysis and discussion. In
addition, we find that ISO~138 has IRS excess emission inconsistent with the ISOCAM  photometry
\citep{persi00,lopez04}, but fully consistent with  the IRAC 8\,\micron{} by \citet{luhman08}, see also Fig.~\ref{fig:spec2}.   The IRS and the IRAC fluxes at
8\,\micron{} are about eight times lower than the ISOCAM flux at 6.7\,\micron . We suspect  that the
source 2MASS~11082238-7730277 at  $\sim$18\arcsec{} from ISO~138 has contaminated the ISOCAM estimates
by \citet{lopez04}. 

The LL1 observations detected all disks in our sample.  
The low LL1 fluxes confirm the previous suspicion that most very low mass objects have flatter disk structures \citep{apai05}. However, given the low S/N these observations do not further our understanding of the dust composition and will not be discussed in the rest of the paper.

We find that 14 out of 17 cool star disks have 10\,\micron{} silicate emission features. These, combined with the sun-like stars and the Herbig Ae/Be stars as described in Sect.~\ref{res:silicates}, make a sample of 72 disks where the dust properties can be studied as a function of stellar mass/luminosity. To compute the strength of the 10\micron{} features we followed the procedure described in \citet{pas08}. The continuum below the feature is estimated as follows. For the sources that have IRS spectra covering the wavelength region between 5.3-14.2\,\micron{} we fit a third-order polynomial between 6 and 8\,\micron{} and between 12 and 14\,\micron.
These sources comprise all the sun-like stars and the following cool stars: ESO-H$\alpha$559, ISO~79,  CHSM~9484, ISO~147, T37, ISO~165, ISO~252, and Hn~13. For sources that have a smaller wavelength coverage we fit a first order polynomial between the minimum wavelength and 8.2\,\micron{} and between 12\,\micron{} and the maximum wavelength. The sources belonging to the last sample are all the Herbig Ae/Be disks and the following cool stars: Cha~H$\alpha$1, Cha~H$\alpha$9, Cha~H$\alpha$2, ISO~138, Cha~H$\alpha$6, and ISO~217.

We report the detection of the $\nu_5$ vibration-rotation band of acetylene (\ctht) around 13.7\,\micron{} and of the $\nu_2$ band of hydrogen cyanide (HCN) around 14.0\,\micron{} toward at least 20 disks in the combined sample of sun-like and cool stars (Sect.~\ref{res:gas}). 
We identify the $\nu_5$ and $\nu_2$ bands of \ctht{} and HCN  using the following procedure. First, we fit a linear continuum in a wavelength region outside the expected emission bands (between 13.4--13.60\micron{} and 14.15--14.3\micron{}). Then, we integrate the flux of the continuum-subtracted spectra within the wavelength region of the low-resolution spectrum where the emission bands are expected to be  (13.6--13.81\micron{} for \ctht{} and 13.81--14.15\micron{} for HCN). The uncertainty on the measured band flux is dominated by two errors: a) the error on the continuum subtraction, which we compute as the standard deviation of the pixels outside the emission feature in the continuum-subtracted spectrum; and, b) the propagation of the measurement errors at each pixel. We classify emission bands as ``firm'' detections or ``possible'' detections, based on the S/N values derived above. We successfully confirmed this classification by individually inspecting every spectra by eye. In the case of the cool star sample ‘possible’ detections have S/N between 2 and 5 while all ‘firm’ detections have a S/N $> 7$. In the case of the 
sun-like sample the S/N estimate is more uncertain for two reasons: i) the fit of the continuum is less certain because the continuum is less flat than that from cool star spectra; ii) the errors at each wavelength are more uncertain because the 
sun-like stars have only 4 exposures per nod position (for comparison cool stars have
more than 10 exposures per nod). Most firm identifications have S/N greater than 3 in the
sun-like star
sample  (possible identifications have S/N slightly lower than 3). Five out of 17 cool star disks have
``firm'' detections of \ctht, other 4 sources have possible detections. Only 1 cool star shows a ``possible'' detection of the HCN rovibrational emission band. Among the 44 
sun-like stars, there are 4 ``firm'' and 3 more ``possible'' detections of \ctht, while there are 13 ``firm'' detections and 7 more ``possible'' detections of HCN (see also Table~\ref{T:det_stat}).

\subsection{Weak Silicate Emission Features in Cool Star Disks}\label{res:silicates}
Grains from the interstellar medium (ISM) provide the initial solids that comprise forming circumstellar disks.
These grains are likely a mixture of silicates and carbon with a range in size from a few tens of
\AA{} up to $\sim$0.2\,\micron{} (see, e.g. \citealt{draine03,gail09}). The 10\,\micron{} absorption features
toward several lines of sight show that most of the silcates are amorphous and crystalline material
constitutes less than 2\% by mass \citep{kemper05}.  These primordial grains undergo major chemical and physical changes during and after the formation of circumstellar disks, some of which are reflected
in the 10\,\micron{} silicate emission feature. Silicate emission features from our targets
show a considerable diversity  ranging from the strong, sharp feature of Cha~H$\alpha\,1$ to the 
almost featureless spectra of J11082570, J11084952, and J11112249. 

The qualitative spectral comparison in Fig.~\ref{fig:featcomp} shows that
dust in the disks of cool stars is more processed than it is in the ISM (see also \citealt{apai05}). 
 In fact, these disks have a broad,
flat-topped 10\,\micron{} emission band, indicative  of grains larger
than the 0.1\,\micron{} grains that dominate the ISM. 
In addition, most spectra show peaks at  9.4 and 11.3\,\micron{} from
crystalline silicates similar to those identified in spectra of 
comets \citep[e.g.,][]{wooden07}.  

Van~Boekel et al.~(2003) first identified a correlation between the shape
and the strength of the 10\,\micron{} silicate emission features from disks around intermediate-mass stars (Herbig Ae/Be stars). This
correlation was interpreted in terms of dust evolution and is often used  to
characterize the amount of ``primitive'' versus ``processed'' dust in disk
atmospheres.  We have re--analyzed the Herbig Ae/Be sample from
\citet{vanboekel05} and expanded the analysis to disks around young 
sun-like stars (from the sample studied in \citealt{pas08}) and cool stars (from this work). In Fig.~\ref{fig:ss},
we show the flux ratio of the continuum-subtracted spectra at 11.3 and
9.8\,\micron{} versus the peak-over-continuum
flux density within the 10\,\micron{} emission feature.  Features dominated by  ``primitive'' dust
(sub-micron amorphous grains like those in the ISM) have low 11.3/9.8 ratios and high peak-over-continuum fluxes. In contrast,
features dominated by ``processed'' dust (micron-sized grains and crystals)
are characterized by high 11.3/9.8 ratios and low peak-over-continuum fluxes. This
figure confirms the trend we reported in \citet{apai05}: disks of cool stars have silicate emission features dominated by ``processed'' dust, the
extent of processing is typically higher than in disks around
sun-like and Herbig Ae/Be stars. Recently, \citet{ks06}
compared the strength of the silicate emission feature from disks around 13 Herbig Ae/Be stars to that 
from disks around 15 M stars and found a probability as large as $\sim$15\% that these two
groups are drawn from the same parent population. 
We applied the Kolmogorov-Smirnov (K-S) test to our much larger sample of  Herbig Ae/Be, sun-like and cool stars. 
We find that the feature strengths from  the Herbig Ae/Be and sun-like star disks and 
from the sun-like star and cool star disks are also consistent with being drawn 
from the same parent population (K-S probabilities of 6 and 7\% respectively). 
On the other hand, the samples of cool star and Herbig Ae/Be disks statistically 
differ in the {\it strength} of their 10\,\micron{} emission features (K-S probability of 0.6\%), with cool star disks having weaker features than Herbig Ae/Be disks. The {\it shape} of the 10\,\micron{} emission feature is not statistically different among the three samples of disks (P(cool:sun-like)=7\%, P(cool:Herbig)=16\%, P(sun-like:Herbig)=6\%). Note that the mean value of the peak-over-continuum becomes larger going from the disks of cool stars, through the sun-like stars, to the Herbig Ae/Be stars (see Fig.~\ref{fig:ss}).
In Sect.~\ref{sect:discuss_dust} we will discuss a scenario that can explain the observed trend between the strength of the silicate features and the stellar mass/luminosity.

\subsection{Low HCN/\ctht Flux Ratios in Cool Star Disks}\label{res:gas}
\begin{figure}
\includegraphics[angle=0,scale=1.0]{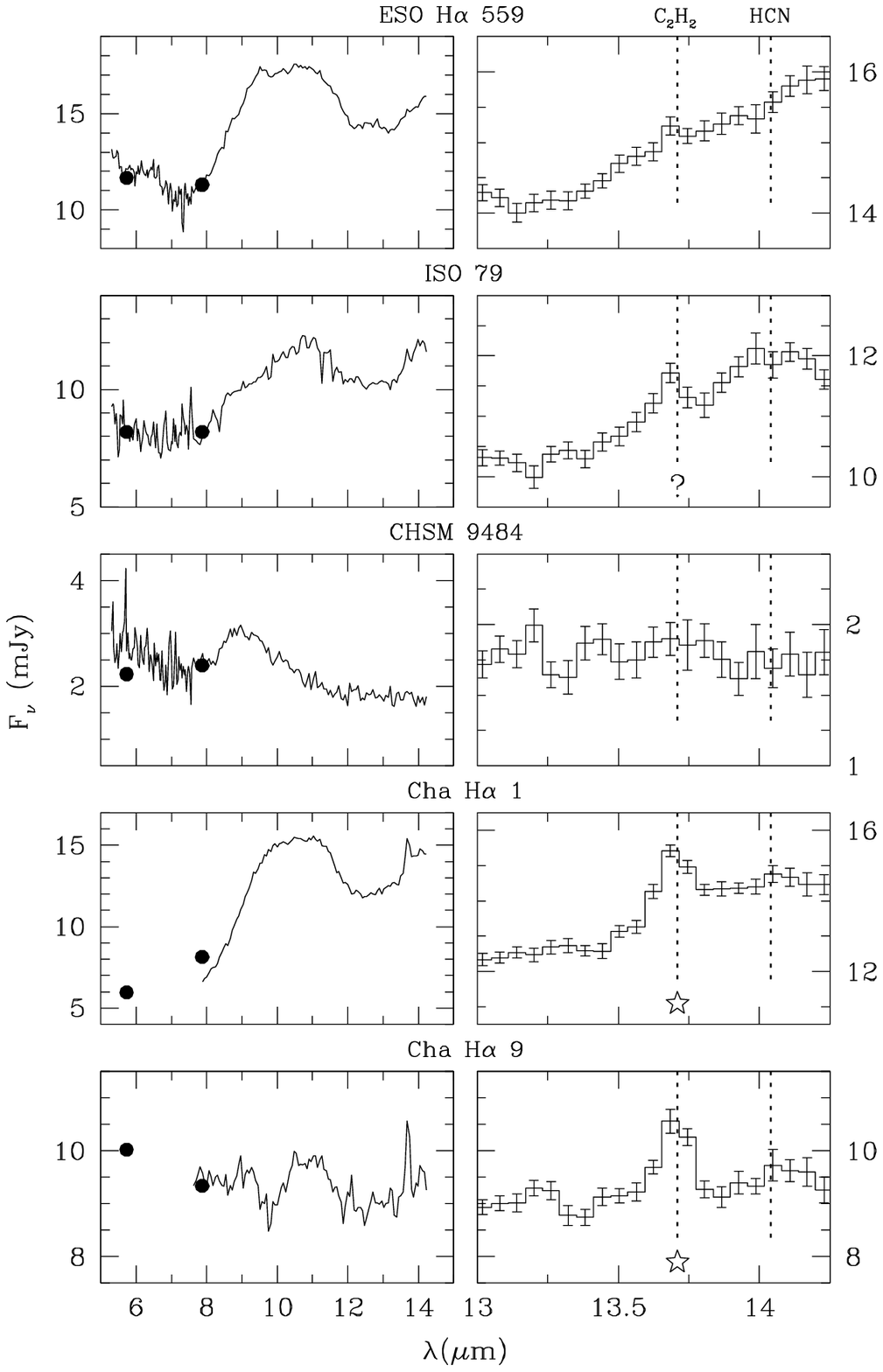}
\caption{Left Panels: 
IRS low-resolution spectra for the cool star sample. 
The IRAC 5.8 and 8\,\micron{} fluxes from \citet{luhman08} are over-plotted as filled black circles (when multiple observations are available we plot the mean flux values). 
Right Panels: 
Expanded view of the brown dwarf spectra around the C$_2$H$_2$ and HCN emission bands. A star indicates a ``firm'' detection while a question mark indicates a ``possible'' detection.
 \label{fig:spec1}}
\end{figure}

\begin{figure}
\includegraphics[angle=0,scale=1.0]{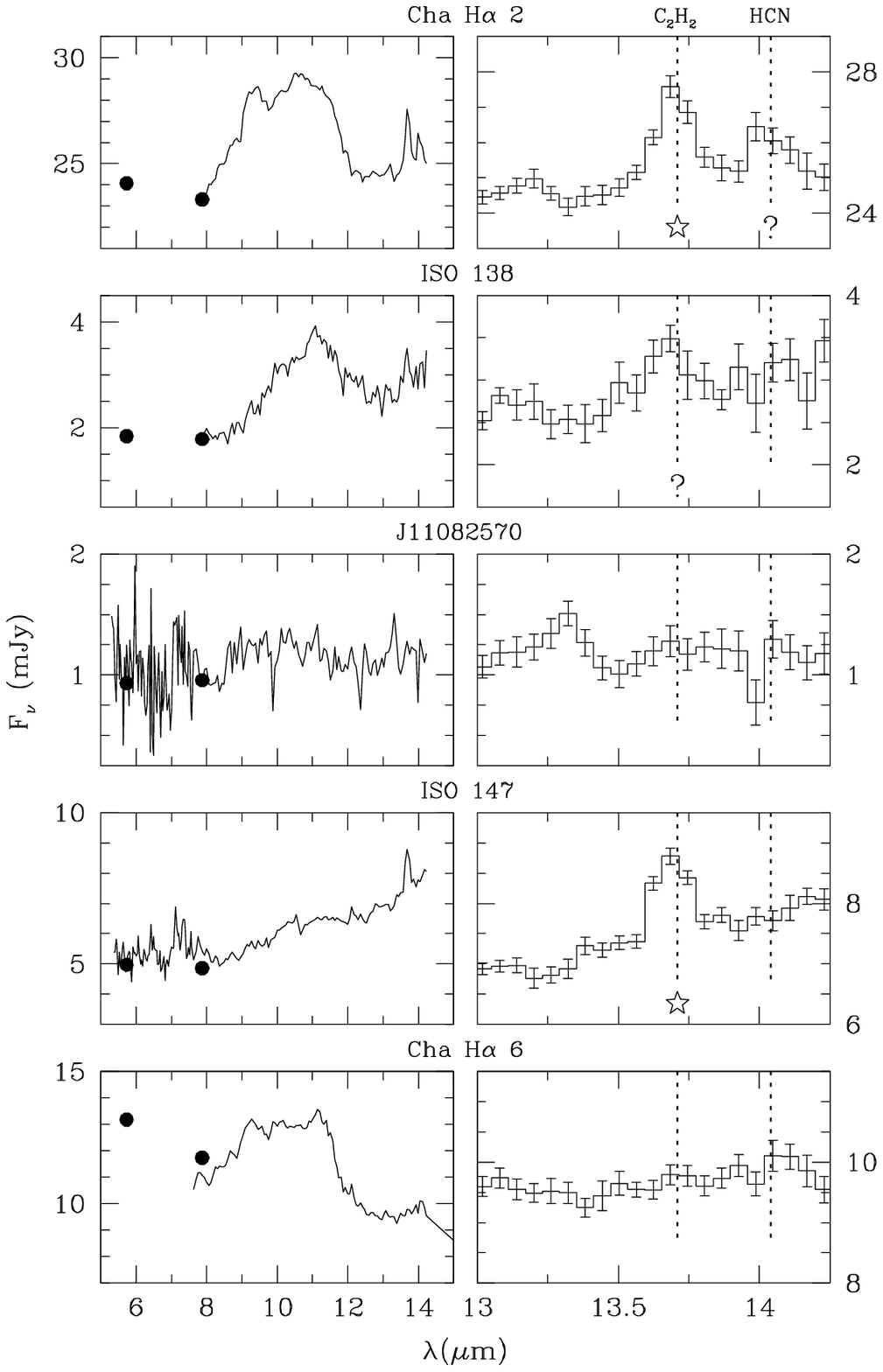}
\caption{Same as Fig.~\ref{fig:spec1}.\label{fig:spec2}}
\end{figure}

\begin{figure}
\includegraphics[angle=0,scale=1.0]{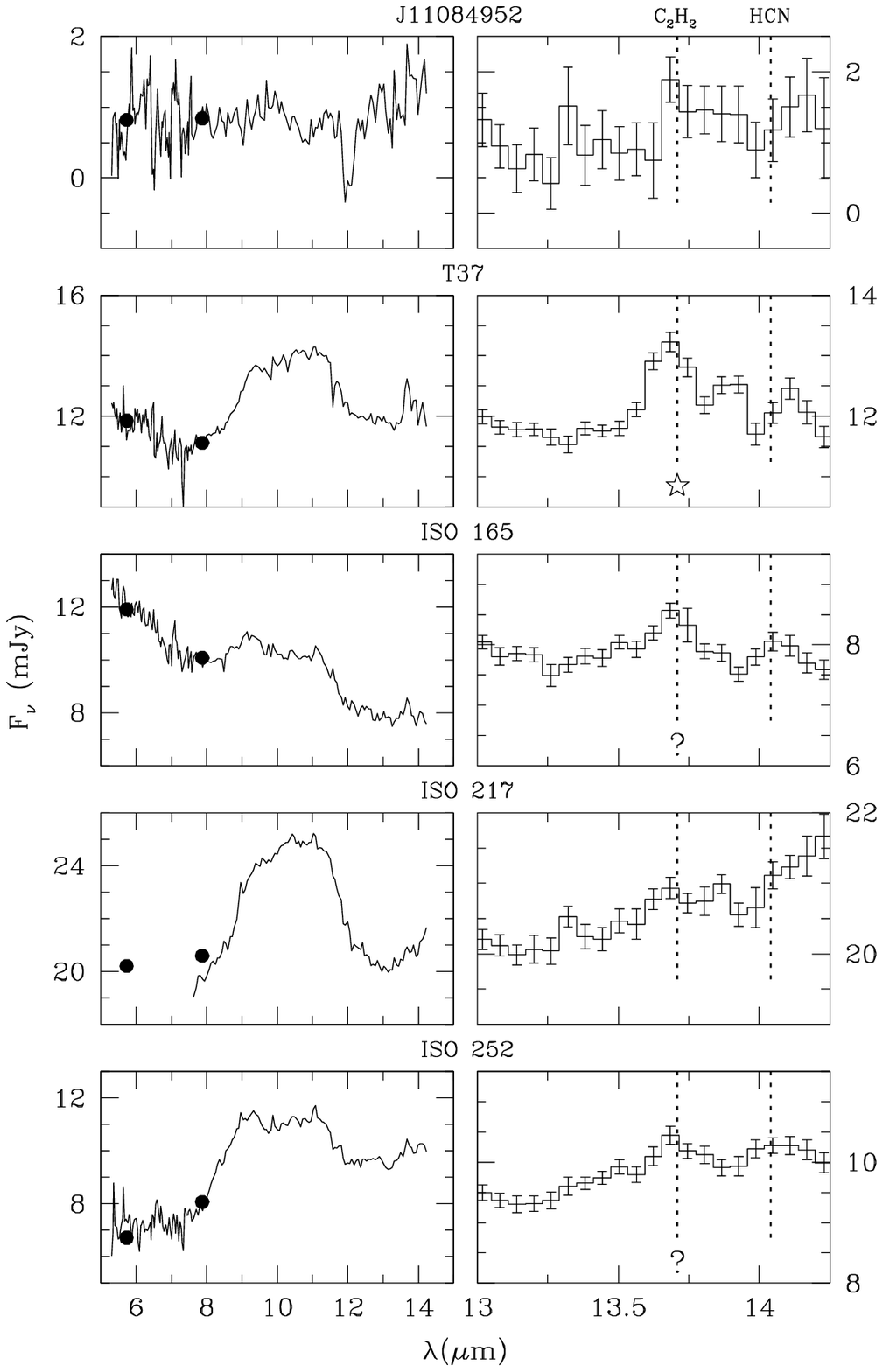}
\caption{Same as Fig.~\ref{fig:spec1}.\label{fig:spec3}}
\end{figure}

\begin{figure}
\includegraphics[angle=0,scale=1.0]{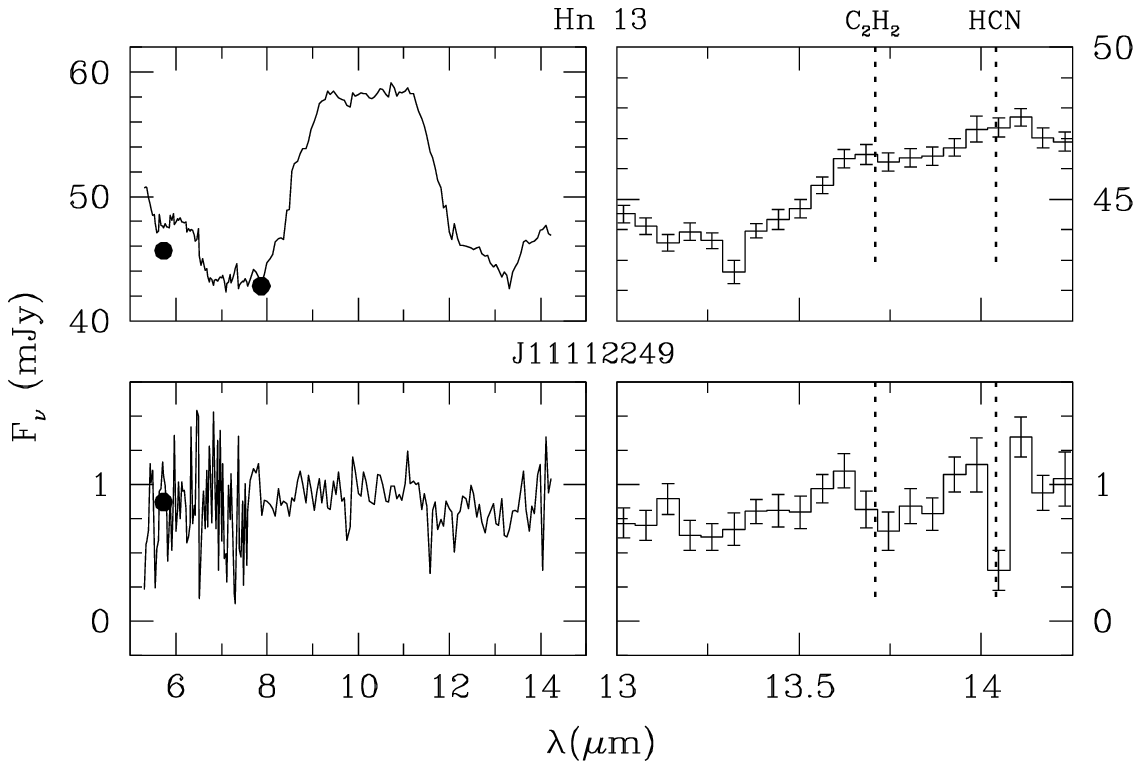}
\caption{Same as Fig.~\ref{fig:spec1}.\label{fig:spec4}}
\end{figure}

\begin{figure}
\includegraphics[scale=0.5]{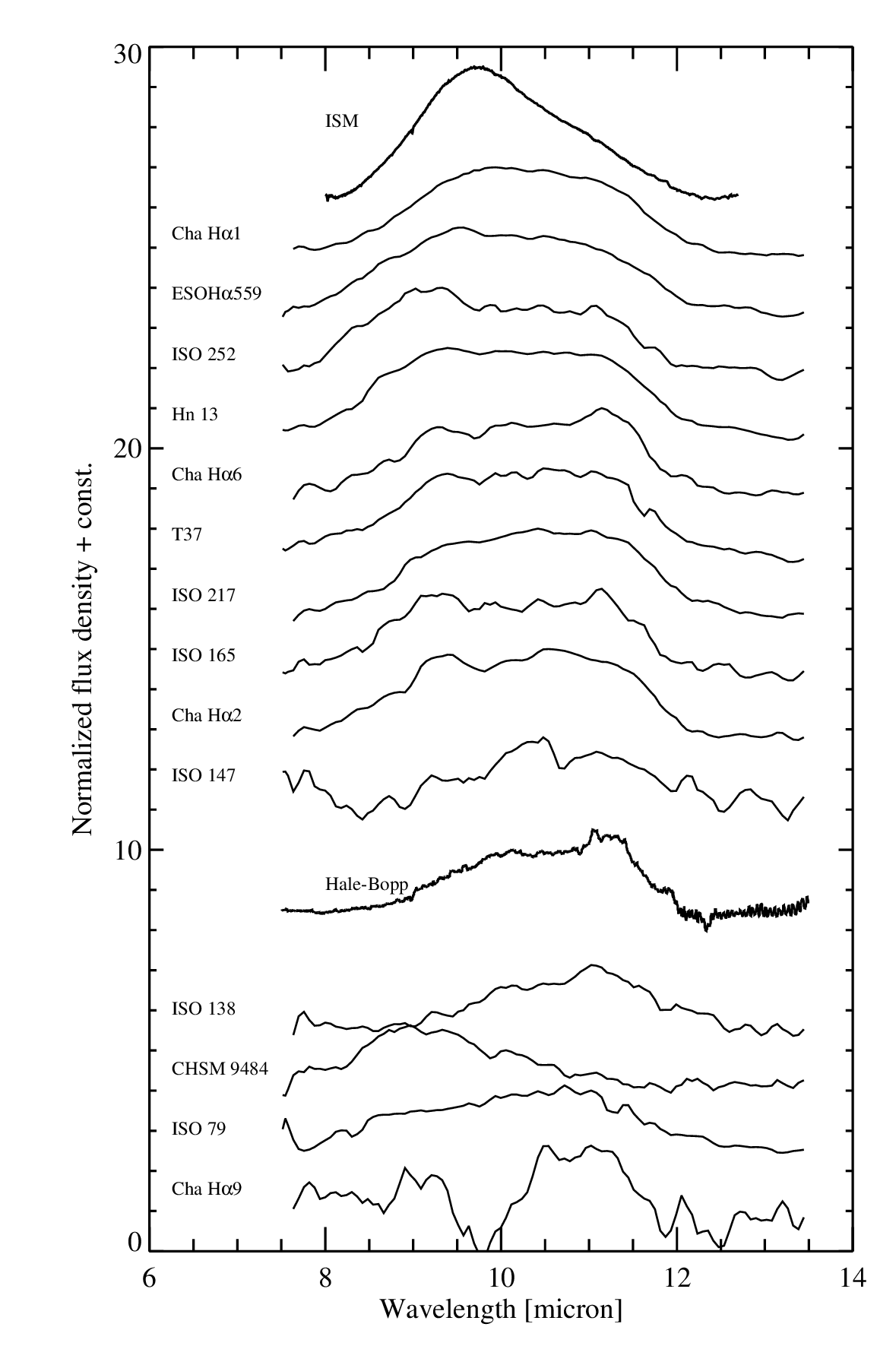}
\caption{Qualitative comparison of the 10\,\micron{} silicate emission features from cool star disks.
Spectra have been continuum-subtracted and normalized to the peak emission.
For comparison the spectra of the amorphous silicate-dominated ISM \citep{kemper05} and the 
crystalline-rich comet Hale-Bopp \citep{crovisier97} are also shown.
\label{fig:featcomp}}
\end{figure}

\begin{figure}
\includegraphics[scale=0.5,angle=90]{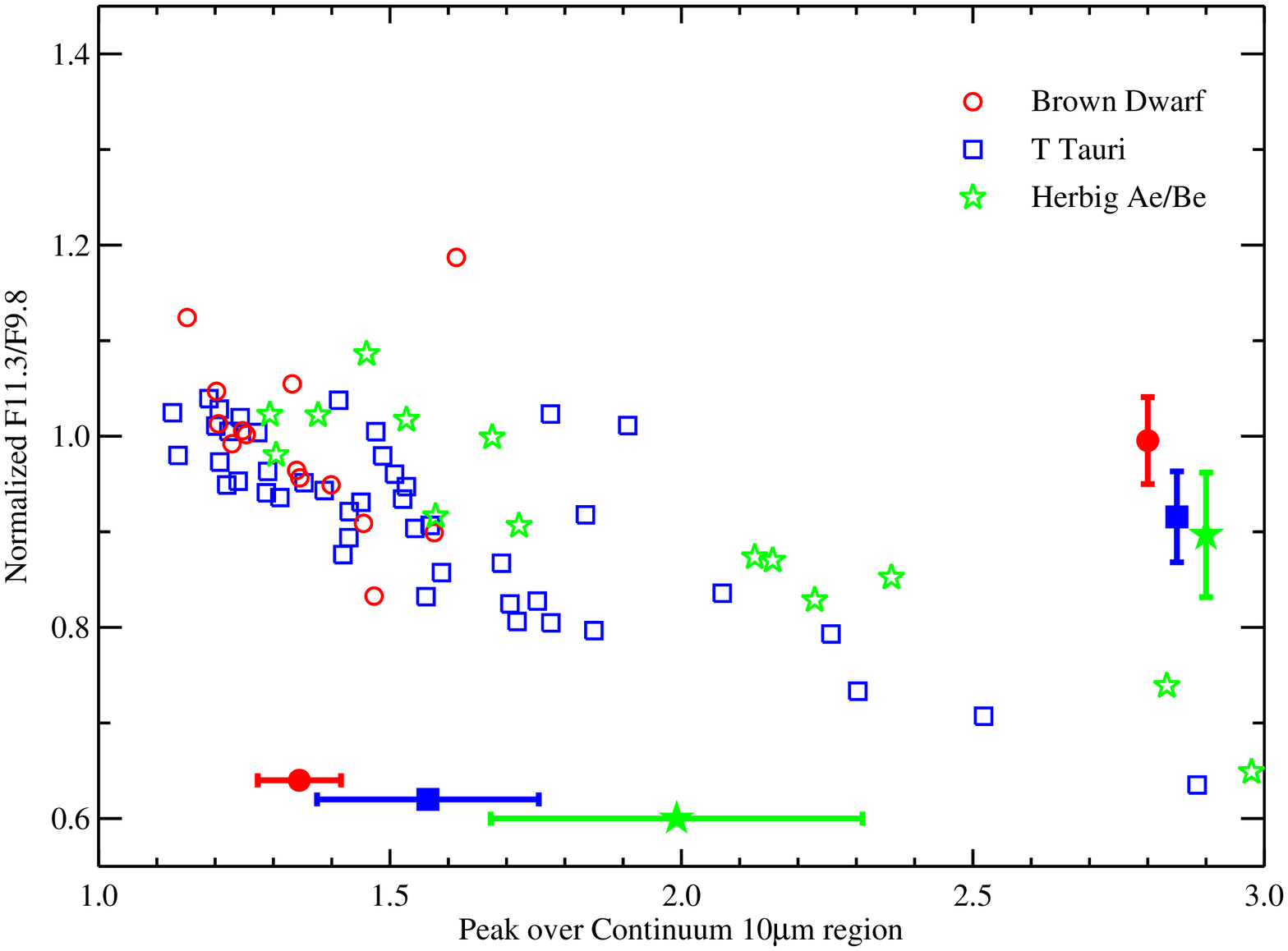}
\caption{Shapes and strengths of the 10\,\micron{} emission features for the samples of Herbig
Ae/Be stars (green stars), sun-like/T~Tauri stars (blue squares), and cool stars/brown dwarfs (red circles). On the y--axis we plot the ratios of the normalized flux at 11.3\,\micron{} over 9.8\,\micron{}, which are proxies for the degree of crystallinity. The x--axis gives the peak over continuum in the 10\,\micron{} region of normalized spectra, which indicates the amount of grain growth. The filled symbols with errorbars show the mean values and the 1$\sigma$ standard deviations of the peak-over-continuum and of the F11.3/F9.8 for the three groups of disks. Cool star disks have statistically weaker 10\,\micron{} features than Herbig Ae/Be disks.\label{fig:ss}}
\end{figure}

\begin{figure}
\includegraphics[scale=1.]{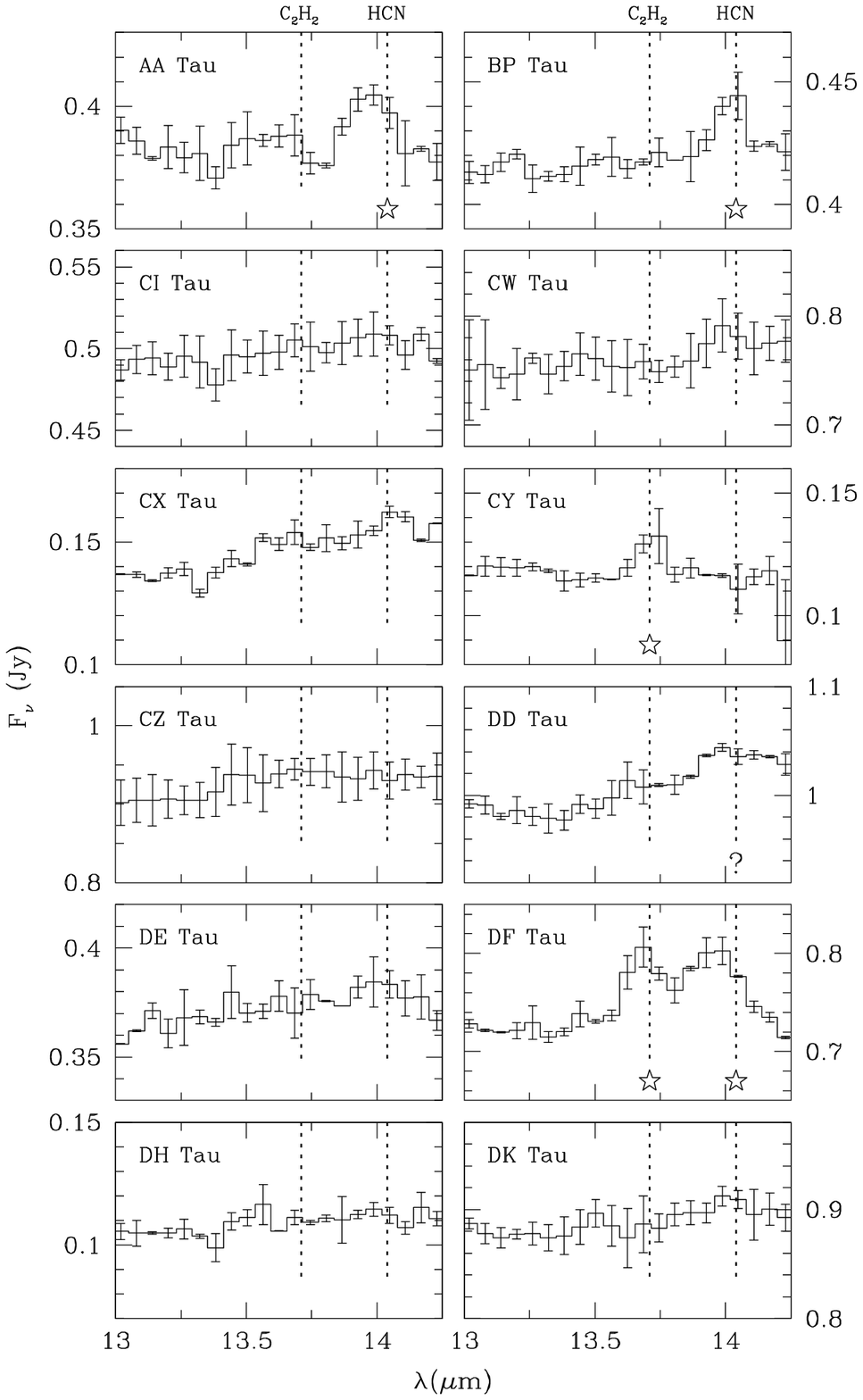}
\caption{Expanded view of the sun-like spectra around the C$_2$H$_2$ and HCN emission bands. A star indicates a ``firm'' detection while a question mark indicates a ``possible'' detection. The full spectra are presented in \citet{pas08}.
\label{fig:spec5}}
\end{figure}

\begin{figure}
\includegraphics[scale=1.]{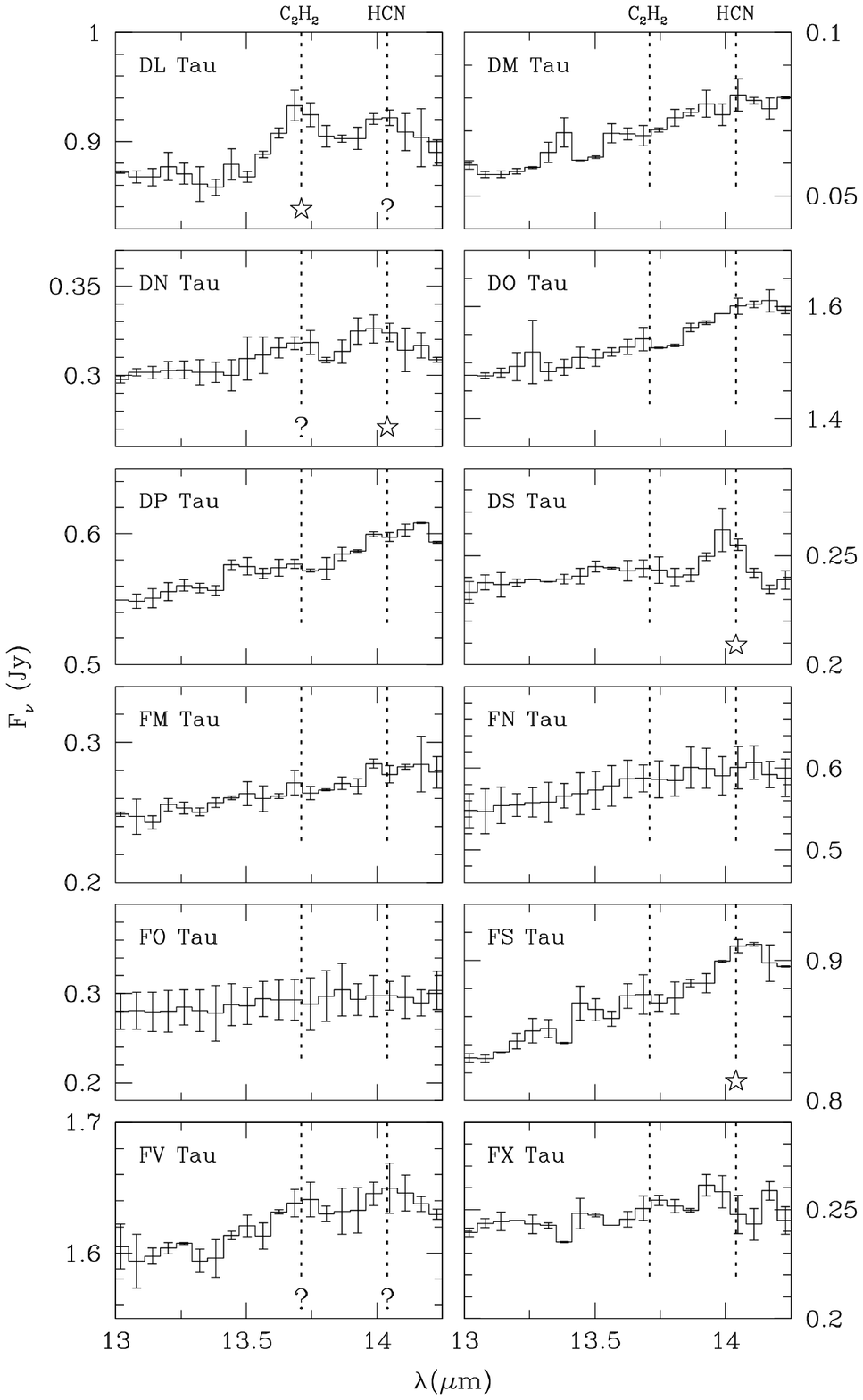}
\caption{Same as Fig.~\ref{fig:spec5}. \label{fig:spec6}}
\end{figure}

\begin{figure}
\centering
\includegraphics[scale=1.]{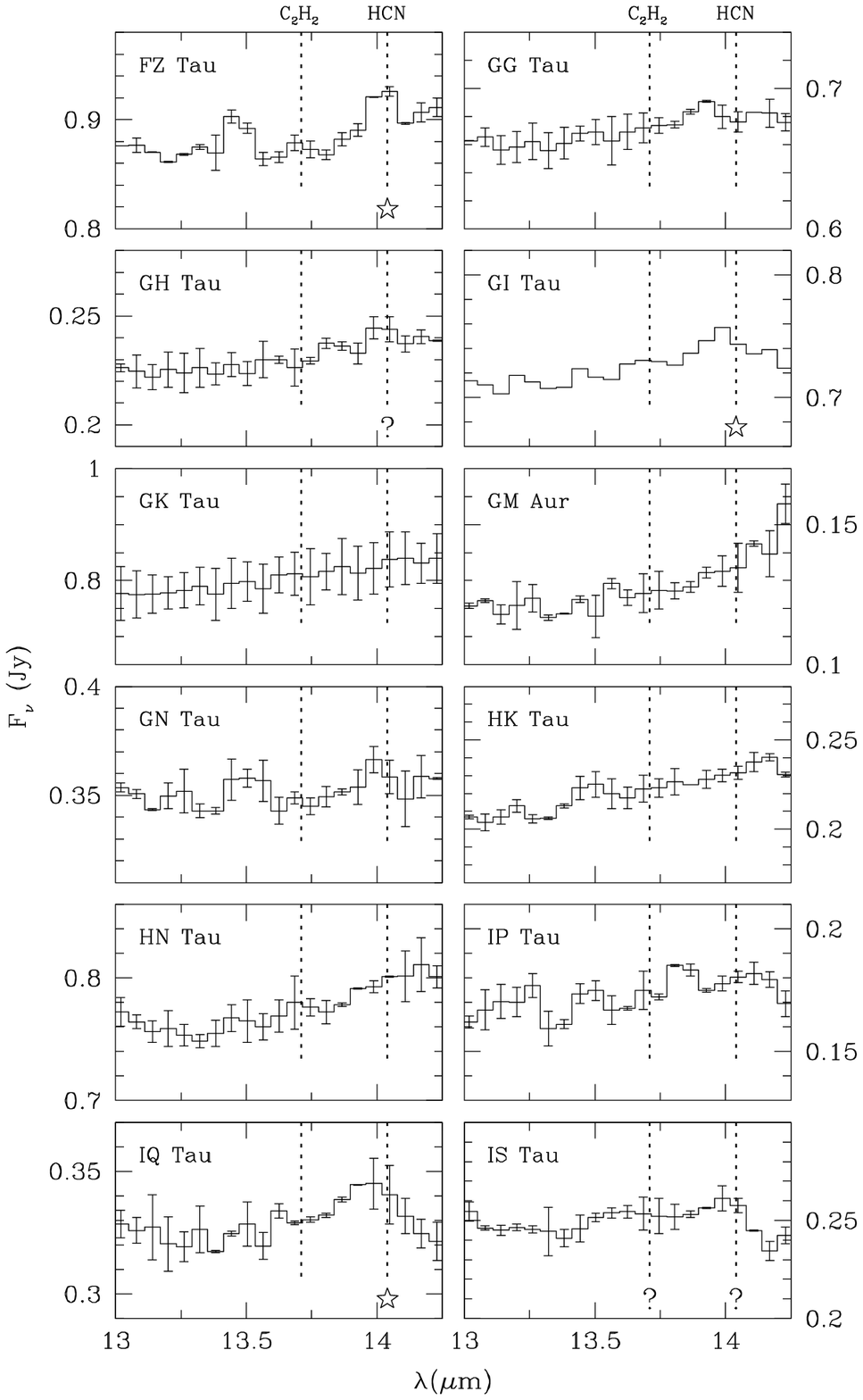}
\caption{Same as Fig.~\ref{fig:spec5}. \label{fig:spec7}}
\end{figure}

\begin{figure}
\includegraphics[scale=1.0]{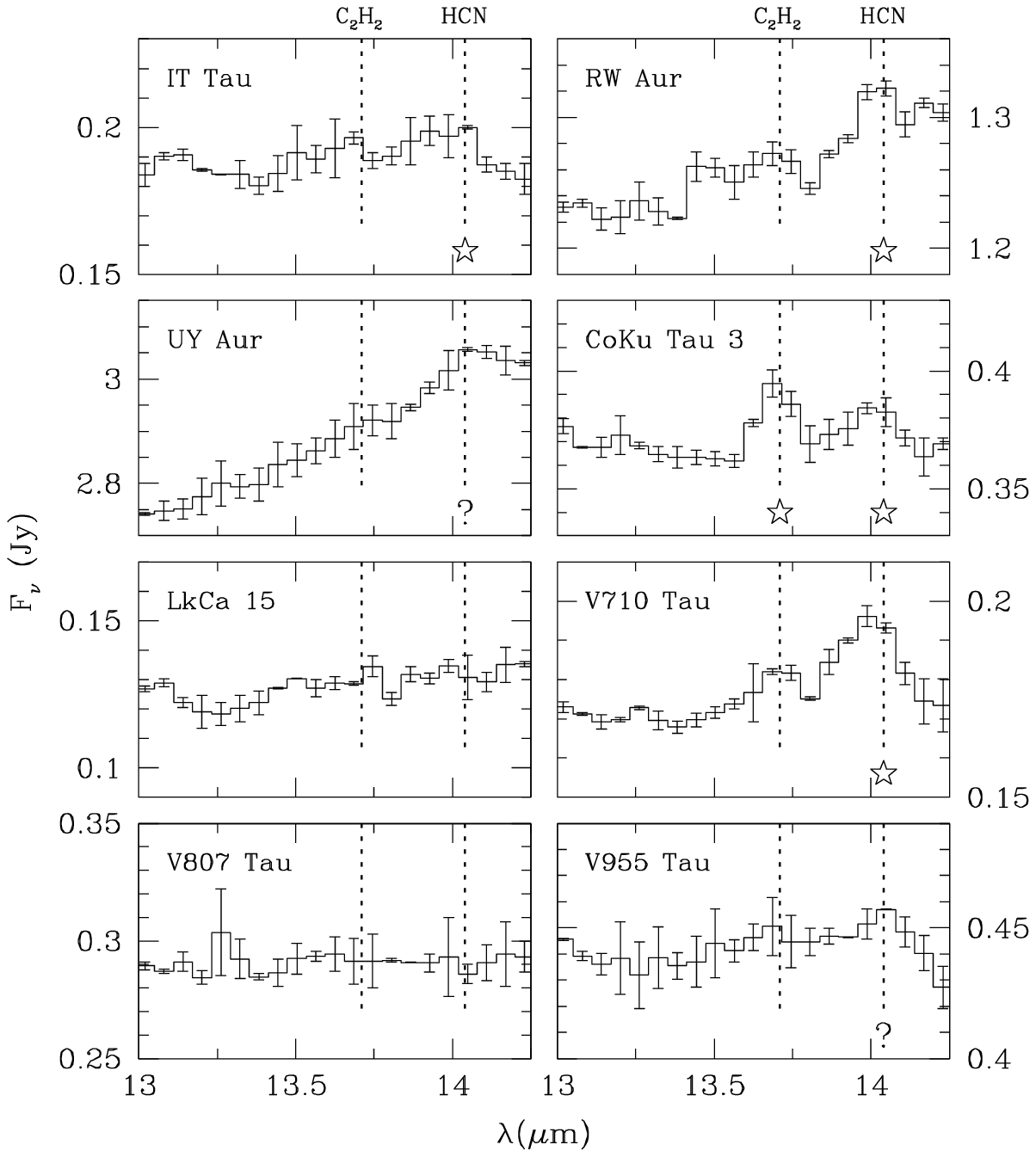}
\caption{Same as Fig.~\ref{fig:spec5}. \label{fig:spec8}}
\end{figure}

The most exciting result of our survey has been the discovery of the vibration-rotation bands of \ctht{} and HCN in at least 5 out of 17 cool stars with disks (Figs.~\ref{fig:spec1}, \ref{fig:spec2}, \ref{fig:spec3}, \ref{fig:spec4}).
No organic molecules have been previously detected toward such very low-mass stars. Motivated by the high detection rate, we have searched for the presence of the \ctht{} and HCN bands in the sun-like star sample. We find that at least 15 out of 44 
sun-like spectra exhibit emission from \ctht{} and/or HCN (Figs.~\ref{fig:spec5}, \ref{fig:spec6}, \ref{fig:spec7}, \ref{fig:spec8}). The sample of sun-like stars consists of about an equal number of single and binary stars \citep{pas08}. The detection rate of \ctht{} and HCN among single stars is similar to that among binary stars. Therefore, in the following we will not distinguish between single and binary stars, but rather contrast the 
sun-like star sample to the cool star sample.
	
\begin{figure}
\centering
\includegraphics[scale=0.8]{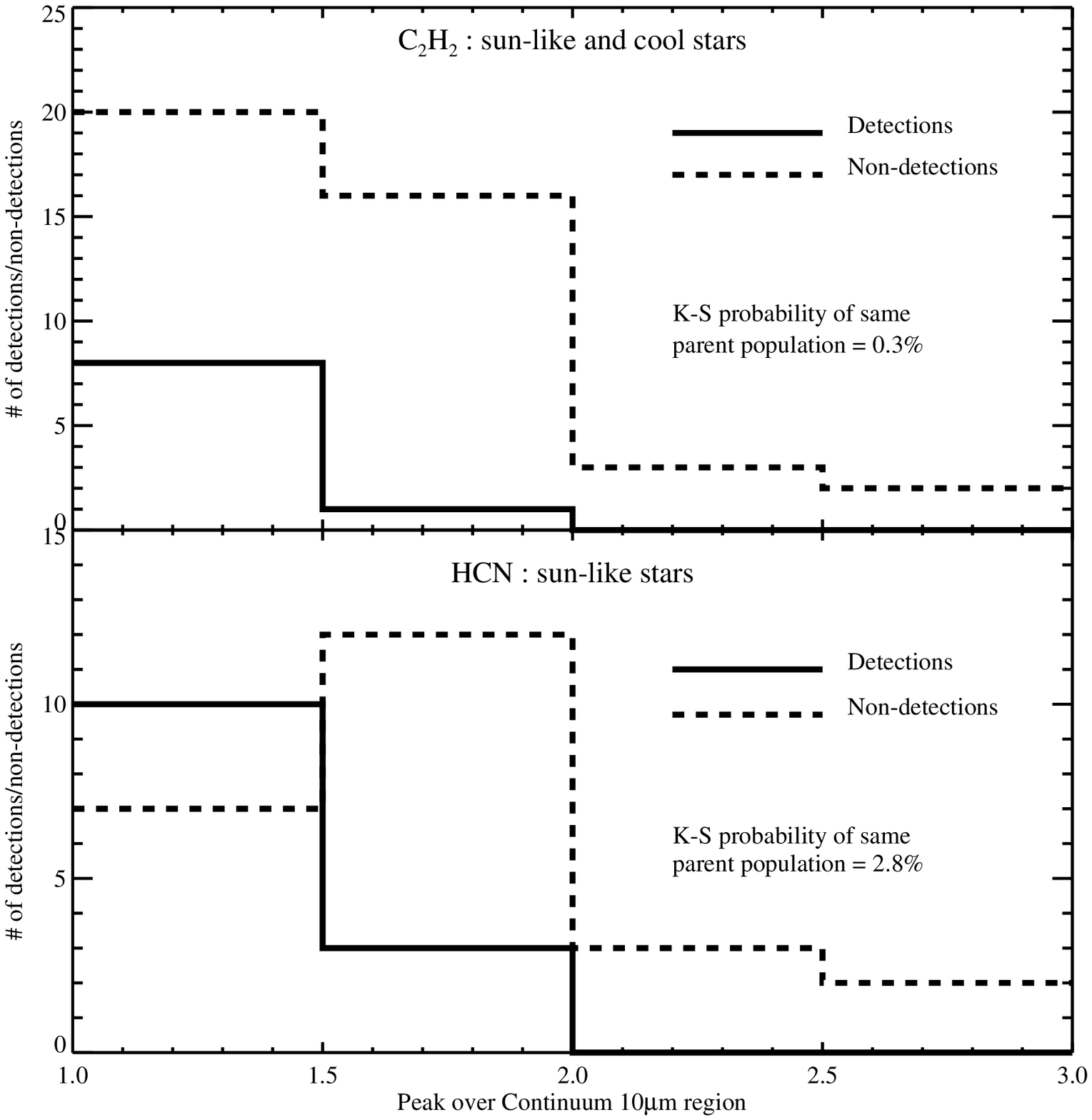}
\caption{Upper Panel: Histogram of the 10\micron{} silicate feature strengths for the detections (solid line) 
and non--detections (dashed line) of the \ctht{} emission band in the combined 
sample of disks around sun-like and cool stars. 
Lower Panel: Histogram of the 10\micron{} feature strengths for the detections (solid line) and non--detections (dashed line) of the HCN emission band in the sample of disks around 
sun-like stars (no cool star spectrum presents a firm HCN detection).
Both panels show  that detections are confined to weak features (peak--over--continuum ratios $< 2$). 
The K-S probabilities that \ctht{} and HCN detections and non--detections are 
drawn from the same parent population are only 0.3 and 2.8\% respectively.
\label{fig:histo_det}}
\end{figure}

\begin{figure}
\centering
\includegraphics[scale=0.8]{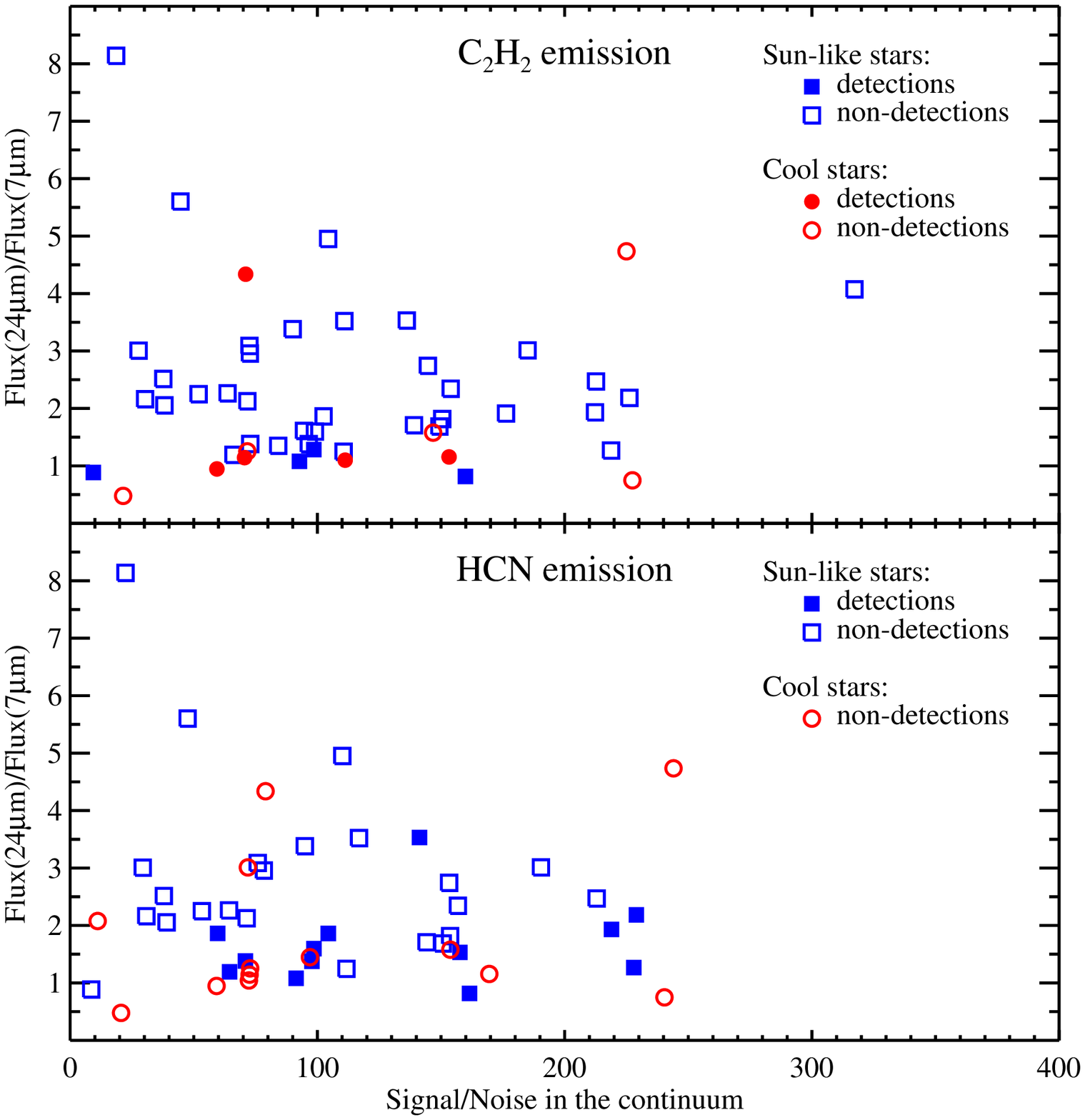}
\caption{Upper Panel : Flux ratios at 24 over 7\,\micron{} versus the signal--to--noise in the continuum close to the rovibrational emission bands of \ctht{} and HCN. Fluxes at 24\,\micron{} are from \citet{pas08} and \citet{luhman08}. Flatter disk structures are indicated by lower flux ratios in the figure. Detections of the \ctht{} band are filled symbols, non--detections are empty symbols (``possible'' detections are omitted). Squares are for sun-like stars, circles for cool stars.
Lower Panel : Same as upper panel but for the HCN emission band. No cool star spectrum presents a ``firm'' HCN detection.
\ctht{} as well as HCN detections are confined to low flux ratios, indicative of flatter disk structures. The figure demonstrates that the pattern of the gas line detections are not primarily set by the signal--to--noise of the observations, thus differences between the two groups of objects compared are not due to different sensitivities.\label{fig:infrared}}
\end{figure}

First, we note that the overall detection rate of \ctht{} and HCN is very similar in the 
sun-like star and cool star samples, with $\sim$30\% of the disks showing emission bands in the Spitzer low-resolution spectra. We searched for any correlation between the presence of gas lines and the dust disk properties, such as the strength of the 10\,\micron{} emission feature and the infrared slope of the source spectral energy distribution\footnote{for this we used the IRS fluxes at 7 and 13\,\micron{}, and the MIPS 24\,\micron{} data from \citet{pas08} and \citet{luhman08}}. Since the 44 sun-like stars were selected to have silicate emission features from optically thin disk atmospheres, we excluded the 3 featureless cool stars from this comparison (J11082570-7716396, J11084952-7638443, J11112249-7745427). Fig.~\ref{fig:histo_det} shows the histograms of the peak--over--continuum ratios of the 10\micron{} silicate emission features for the detections and for the non-detections of the \ctht{} and HCN rovibrational bands. Detections are only present in disks with weak silicate emission features (peak-over-continuum ratios $< 2$). In addition, detections are confined to those disks with evidence for flatter geometries (Fig.~\ref{fig:infrared}). In summary, the disks with gas detections show clear signs of grain growth and dust settling both among the sun-like and the cool star samples. This correlation may be an observational bias. Disks with larger grains and flatter disk geometries have weaker mid-infrared continuum levels making it easier to detect gas emission lines (see, e.g. \citealt{pascucci07}).

When inspecting the detection rates of \ctht{} and HCN separately, we find a striking difference between the samples of sun-like and cool stars. \ctht{} is detected in at least 5 out of 14 cool star spectra with silicate emission features, but at most in 7 out of 44 sun-like stars (only 4 are considered 'firm' detections). In contrast, only the single cool star spectrum of Cha~H$\alpha$2 presents a possible HCN emission band, but at least 13 
sun-like stars have HCN detected (see Table~\ref{T:det_stat}). In addition, the band strengths of \ctht{} and HCN are different in the two samples. Emission from HCN is brighter than that from \ctht{} in most disks around sun-like stars (a notable exception is CY~Tau). In contrast, all disks around cool stars have brighter \ctht{} emission compared to HCN. 
 This trend is well illustrated in Fig.~\ref{fig:plcomb_scaled}, where we plot the median (at each wavelength) of the continuum-subtracted and normalized spectra for the 
sun-like and the cool stars with \ctht{} and/or HCN emission band detections. The flux ratio of HCN over \ctht{} is $\sim$2.9 for the median sun-like star spectrum, while it is only 0.26 for the median cool star spectrum. The only two sources where both \ctht{} and HCN are firmly detected are the sun-like stars DF~Tau and CoKu~Tau~3 with HCN over \ctht{} flux ratios of $\sim$1.3 and 0.7, respectively (see Fig.~\ref{fig:plcomb_scaled}). These two objects alone point to a diversity in the line flux ratios of HCN and \ctht{} in sun-like stars. This diversity should be further investigated with sensitive high-resolution Spitzer spectra. We stress, that the differences in detection frequency and band strength between the sun-like and cool stars are not explained by possible differences in the infrared continuum, its shape, and signal--to--noise ratio among the two samples because the \ctht{} and HCN bands are at adjacent wavelengths. Sect.~\ref{discuss:gaslines} suggests a possible scenario to explain these findings.
											      
\section{Discussion}\label{sect:discussion}

\subsection{Dust Processing, Stellar Luminosity and Disk Turbulence}\label{sect:discuss_dust}
The main result of Sect.~\ref{res:silicates} is the trend between the
stellar mass/luminosity and the strength of the 10\,\micron{} silicate emission
feature with disks around Herbig Ae/Be stars having statistically stronger features
than disks around cooler/lower-mass stars. In other words, the 10\,\micron{} emission features
from disks around low-mass/luminosity stars are dominated by more ``processed'' dust 
than the same features from disks around higher mass/luminosity stars. How can we explain this observational trend? 

The 10\,\micron{} silicate emission feature arises from different regions in disks around stars of different luminosities. \citet{ks07} calculated that it probes radii $\ge$ 0.5--50\,AU in disks around Herbig Ae/Be stars, but only radii $\le$0.001--0.1\,AU in brown dwarf disks. 
This fact may provide an explanation to our observations. The closer the grains are 
to the central star the faster they are depleted due to the higher density and rotational frequency which
enhance the collision rate of dust grains and the velocity at which particles settle to the disk midplane.
In this picture, the grain sizes inferred from the 10\,\micron{} silicate features are not representative for the bulk silicate dust in the disk but rather reflect the different location of the 10\,\micron{} emission zone around stars of different luminosities (see also \citealt{apai05,ks07}).
However, grain growth and dust settling are not the only mechanisms determining the grain size distribution in the disk atmosphere. 

Turbulence may also affect the strength of the 10\,\micron{} emission feature, for instance by  stirring up large grains from the disk midplane. In the frame of the magnetorotational instability model \citep{balbus91}, stellar accretion rates are a good proxy for the disk turbulence \citep{gammie96}.  In Fig.~\ref{fig:Mdotvs10um} we plot the strength of the 10\,\micron{} emission features versus the stellar accretion rates for the sample of sun-like stars. The upper envelope of this plot suggests a trend with high accretion rate systems having large grains and moderate/low-accretion rate systems having small grains. However, the large uncertainty in the stellar accretion rates precludes any firm conclusion. This, in addition to the fact that large grains are seen commonly at all accretion rates, suggests that accretion/turbulence can only play a minor role in determining the strength of the 10\,\micron{} silicate emission feature.


Finally, it should be also noted that disk models including only dust coagulation and settling/vertical mixing produce a too rapid (only 10$^4$\,years) depletion of grains up to 100\,\micron{} in size essentially at all disk radii \citep{dd05}, which is inconsistent with the frequent detection of 10\,\micron{} emission features in Myr-old disks.
Fragmentation of aggregate grains could be the major mechanism replenishing the population of small grains.
\citet{brauer08} implemented aggregate fragmentation in their disk model and show that grain growth and fragmentation can reach an equilibrium for a period of $\sim10^6$  which results in a quasi-stationary grain size distribution. If the average grain size of quasi-stationary grain size distributions decreases with radial distance from the central star then such disk models could explain our observations.

\subsection{HCN under-abundance in Disk Atmospheres of Cool Stars}\label{discuss:gaslines}
We have shown in Sect.~\ref{res:gas} that there is a striking difference in the detection rates
of HCN and \ctht{} between the sun-like and the cool star samples. HCN is often present in 
sun-like star spectra,
but absent from the spectra of cool stars. Similarly, Fig.~\ref{fig:plcomb_scaled} demonstrates that the HCN/\ctht{} flux ratio is much higher toward sun-like stars than toward cool stars.
As explained below, these differences arise from different abundances in the two samples. 


To model the emission from the $\nu_5$ band of \ctht{} and the $\nu_2$ band of
HCN we followed the procedure described in \citet{lahuis00}, which assumes   
rotational and vibrational levels in local thermal equilibrium. 
Using the populations in each level, the optical depths are
calculated assuming a Voigt profile function. The spectrum is also
multiplied at each frequency by the black body emission of gas at the temperature T$_{\rm gas}$ and 
reduced to the resolution of the IRS/SL module using a Gaussian profile.
First, we explored whether our low-resolution spectra can provide any interesting constraints on 
T$_{\rm gas}$. To do that, we have simulated spectra with gas temperatures from 100 to 900\,K
and measured the shift in the band head peak emission. 
The peak emission of the $\nu_2$ band of HCN shifts from 14.02\,\micron{} for T$_{\rm gas}$=100\,K to 13.96\,\micron{} for T$_{\rm gas}$=900\,K.
This shift is about half the spectral resolution of the IRS/SL1 module at 14\,\micron{} and cannot be reliably detected in our spectra. A similar result is reached for the $\nu_5$ band of \ctht . We find that the wavelengths at which the emission of HCN and \ctht{} peak in the observed spectra are in between the predicted peaks for gas at 100\,K and 900\,K. In summary, the resolution of our spectra is not sufficient to measure the temperature of the emitting gas, but it is fully consistent with the emitting gas having temperature between 100\,K and 900\,K.

More excitingly, the HCN/\ctht{} flux ratio informs us on the relative column densities of the two
molecules. We computed synthetic spectra with varying column density ratios of HCN over \ctht{},
calculated the line flux ratios and compared them to the observed line flux ratios in the  sun-like
and cool star samples (see Fig.~\ref{fig:abun_fig}).  
 Absorption spectra of HCN and \ctht{} toward
young embedded massive stars find highly correlated excitation temperatures for these molecules
demonstrating that they probe warm gas at similar temperatures \citep{lahuis00}. We consider here
three temperatures for the HCN and \ctht{} emitting gas: 650\,K, which \citet{carr08} find to fit
best the HCN and \ctht{} band profiles from AA~Tau, 450 and 850\,K as comparison temperatures.
Other important model parameters are the gas column densities and Doppler broaden line widths.
In our calculations we keep the \ctht{} column density fixed to 10$^{16}$\,cm$^{-2}$ while we vary the HCN column density between 0.5-11$\time10^{16}$\,cm$^{-2}$, thus covering the column densities derived by \citet{carr08} for AA~Tau. We also assume a Doppler line broadening of 5\,km/s, greater than the thermal broadening for the considered temperature range. The strongest \ctht{} and HCN emission lines start to become optically thick around 10$^{16}$\,cm$^{-2}$ and 7$\times10^{16}$\,cm$^{-2}$, respectively, for gas temperatures $\ge$400\,K (see Fig.~3 from \citealt{lahuis00}).  
Fig.~\ref{fig:abun_fig} shows that, regardless of the gas temperature, HCN/\ctht{} line flux ratios
as high as 3 can be obtained only for large  HCN/\ctht{} column density ratios. 
The line ratio is almost independent from the gas temperature because both the \ctht{} and the HCN lines start to become only marginally thick for the parameters we considered. If we assumed higher column densities for the two molecules or smaller Doppler line broadening, the HCN over \ctht{} line flux ratio would start to saturate faster and even higher HCN/\ctht{} column density ratios would be required to obtain the typical HCN/\ctht{} line ratio of 3 of sun-like stars. In contrast, the low HCN/\ctht{} flux ratio measured in cool stars must originate from a lower column density of HCN in respect to \ctht.
For comparison, \citet{lahuis00} and \citet{boonman03} find that HCN is a
factor of a few more abundant than \ctht{} in hot molecular cores.  How can we explain the HCN
under-abundance in the disk atmospheres of cool stars? 

Here we propose a possible explanation. If HCN is formed in the warm disk atmosphere then its abundance may be limited by the availability of gas--phase atomic N (e.g. \citealt{agundez08}). Atomic N is primarily produced via the photo--dissociation of N$_2$, thought to be the most abundant nitrogen-bearing molecule in the interstellar medium \citep{bergin95}. Thus, the HCN abundance may be limited by $\Phi_{\rm uv}$, the flux of UV--photons with $\lambda < 110$ nm, required to dissociate N$_2$\footnote{While the dissociation of NH$_3$ could also provide free N, \citet{vandishoeck08} show that photodissociation rates for this molecule are  insensitive to the radiation field for temperatures between 10,000 and 4,000\,K}. For young stars $\Phi_{\rm uv}$ is dominated by emission of hot ($\sim$10,000\,K) gas accreting onto the stars \citep{matsuyama03, herczeg04}. A simple estimate for the accretion luminosities of the two samples suggests that their  $\Phi_{\rm uv}$ differ by more than 3 orders of magnitude, although the stellar temperatures differ by only 30\% (Table~\ref{T:det_stat}). If the HCN production from N is indeed driven by the UV-–photochemistry, the large difference in the UV flux of 
sun-like and cool stars may naturally explain our observations. An important consequence of this scenario is the general under--abundance of molecular species produced through UV chemistry and an over--abundance of molecules with large binding energies in cool star disks. With efficient vertical and radial mixing (e.g., \citealt{ciesla07}), the composition of the disk atmosphere may be representative of the overall disk composition. Indeed, turbulent mixing seems to be necessary to explain the high abundances of  warm HCN and \ctht{} in the only single disk atmosphere where they were detected previously \citep{carr08}. 
Detailed dust and gas disk modeling is required to understand whether the UV-photochemistry in the disk atmosphere can radically affect the organics in the disk midplane.

An alternative, though perhaps less likely explanation is that our observations are tracing cometary or interstellar ice evaporation, in which case the organics in the disk midplane may not be different. \citet{disanti08} have shown that Oort cloud comets have an HCN/\ctht{} ratio near unity, but Jupiter family comets have  less \ctht. Because the HCN abundance relative to water is high by interstellar standards in Oort cloud comets, HCN seem to form efficiently in the outer disk, perhaps via surface chemical reactions on icy dust grains \citep{charnley92}. Nonetheless, the evaporation of cometary ices alone cannot reproduce our results, unless \ctht{} is preferentially destroyed in 
the disk atmospheres of sun-like stars. \ctht{} has a dissociation cross-section that is at least an order of magnitude higher than HCN for photons with $\lambda$ between 1300--2000\,\AA{} (\citealt{lee84}). Because 
sun-like stars have more flux than cool stars at these wavelengths, \ctht{} could be more easily destroyed in the disk atmosphere of sun-like stars. However, it is interesting to consider how photodissociation rates depend on the stellar radiation field (Table~2 from \citealt{vandishoeck08}). Going from a stellar radiation field produced by a star of 10,000\,K to that of a 4,000\,K star the \ctht{} photodissociation rate decreases only by a factor of 9, that of HCN by a factor of 68, while that of N$_2$ by more than 4 orders of magnitudes. Based on this dependence on the temperature of the radiation field, it seems more likely that the large difference in HCN/\ctht{} abundance is caused by more efficient dissociation of N$_2$ in disk atmospheres of accreting sun-like stars. 

\subsection{On the Organics Available During Planet Formation}
HCN is the key building block of many complex organic compounds. Under reducing conditions the polimerization of five HCN molecules can form adenine \citep{oro60}, one of the two purine nucleobases at the core of the replication system of all terrestrial organisms. But even if the early Earth atmosphere was reducing, the synthesis of adenine through HCN polymerization under terrestrial conditions remains controversial 
\citep{shapiro95} and its delivery through cometary or meteoritic impacts has been proposed as a likely alternative \citep{chyba92}. Adenine synthesis likely occurs in extraterrestrial environments \citep{stoks81,kissel87}, consistent with the observations of abundant HCN in the interstellar space \citep{hirota98}, in comets \citep{magee02} and recently in disks (this work, \citealt{lahuis06,gibb07,najita07}). In addition, the availability of simple organic molecules such as HCN has a fundamental impact on the chemical networks in protoplanetry disks \citep{semenov04} and thus on the compounds available during planet formation. 

Our observations show that HCN is under-abundant in the disk atmosphere of cool stars. With efficient mixing in the disk, this finding may reflect an overall deficiency of HCN in disks around cool stars. In this case, it is possible that the organics in the planet forming regions of cool stars, as well as the organics possibly delivered post-formation, will differ from those around sun-like stars. If exogenous HCN has played a key role in the synthesis of prebiotic molecules on Earth as proposed, then prebiotic chemistry may unfold differently on planets around cool stars.

\section{Summary}
Using Spitzer/IRS we compared the gas and dust properties of 17 disks around cool stars/brown dwarfs 
to 58 disks around sun-like and intermediate-mass stars. The cool stars have spectral types later than M5, the sun-like stars between K1 and M5, and the intermediate-mass stars have spectral types earlier than A0.
 As the first large comparison, 
we found important and surprising differences between these samples.
Our results can be summarized as follows:
\begin{enumerate}
\item We identify 14 cool star disks with 10\,\micron{} silicate emission features. Features are broad and weak suggesting that grains at least a few-microns in size are present in the disk atmospheres. 
Most of the cool star spectra also show prominent crystalline silicate features.

\item The sample of disks around cool and intermediate-mass stars statistically differ in the strength of their 10\,\micron{} emission features, with cool stars having much weaker continuum-subtracted features than intermediate-mass stars. The observed difference demonstrates that the dust populations at the observed disk radii are different in the two samples.  


\item We report the first detections of organic molecules from disks around brown dwarfs. 

\item We find a striking difference in the detection rates of HCN and \ctht{} between the sun-like and the cool star samples, with HCN often present in sun-like stars, but absent from the cool star spectra. Similarly, the HCN/\ctht{} flux ratio of the median sun-like star spectrum is an order of magnitude higher than that of the median cool star spectrum.
We interpret these as evidence for  different abundances of the two molecules in the disk atmosphere of sun-like and cool stars. Efficient photo–dissociation of N$_2$ in the disk atmosphere of sun-like stars may drive the higher production of HCN.
\end{enumerate}

\begin{figure}
\centering
\includegraphics[scale=0.8]{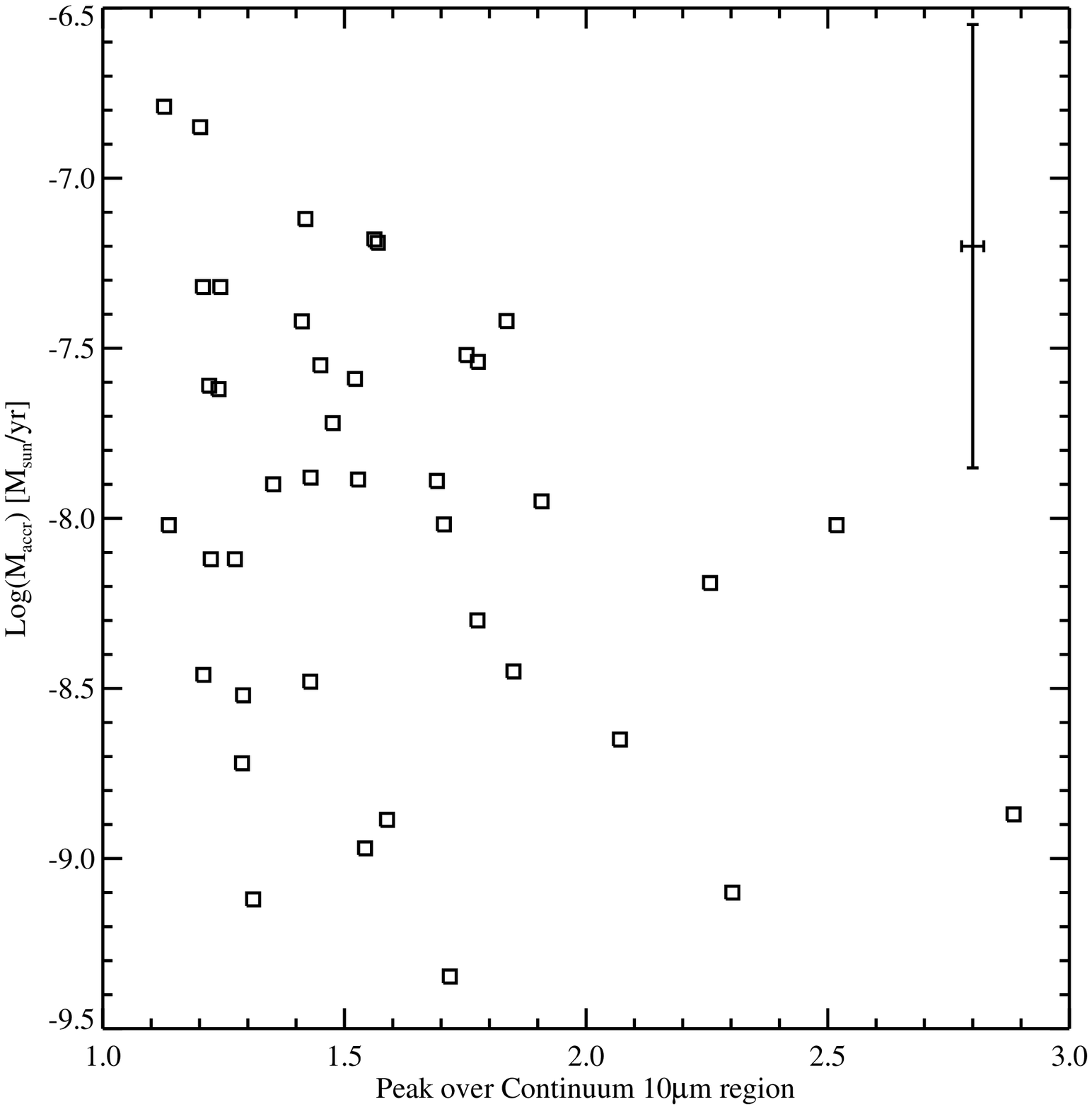}
\caption{Stellar accretion rate (M$_{\rm accr}$) versus 10\,\micron{} feature strength for the sample of sun-like stars.
Accretion rates are from the literature \citep{hartigan95,hartmann98,gullbring98,wb03,najita07} and have been placed on a consistent scale using the factors calculated in \citet{najita07}. 
FM~Tau, CoKu~Tau~3, IT~Tau, and V710~Tau do not have measured accretion rates. 
The errorbar on the peak-over-continuum is from \citet{pas08} while the 
uncertainty in the mass accretion rate is from \citet{najita07}.
There is a 4\% probability that the data are randomly distributed, the negative Kendall rank correlation coefficient  (-0.22) indicates a moderate anti-correlation. Yet, given the large uncertainty in the mass accretion rate we are cautious of concluding this. \label{fig:Mdotvs10um}}
\end{figure}

\begin{figure}
\centering
\includegraphics[scale=0.5,angle=90]{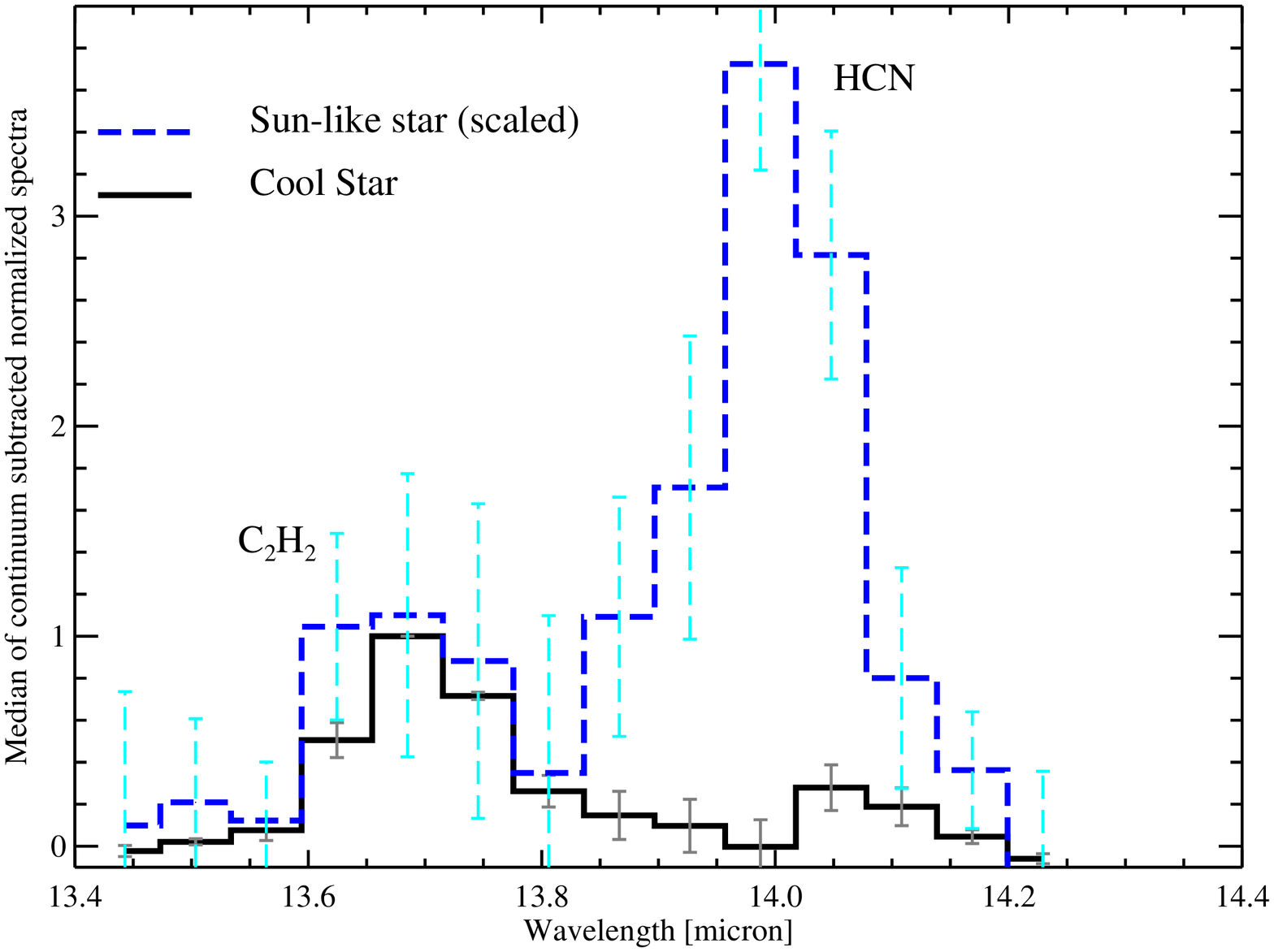}
\caption{Median of continuum-subtracted and normalized spectra for the sun-like star (blue) and the
cool star (black) samples presenting \ctht{} and/or HCN emission bands (5 cool stars and 15 sun-like stars). 
The spectra are normalized to
the peak of emission and scaled to match the \ctht{} emission in the two samples. The errorbars are
the standard deviations of the normalized spectra. If cool stars had the same flux ratio of HCN
versus \ctht{} as the  sun-like stars do, HCN emission would have been easily detected toward
them.\label{fig:plcomb_scaled}}
\end{figure}

\begin{figure}
\centering
\includegraphics[scale=0.5, angle=90]{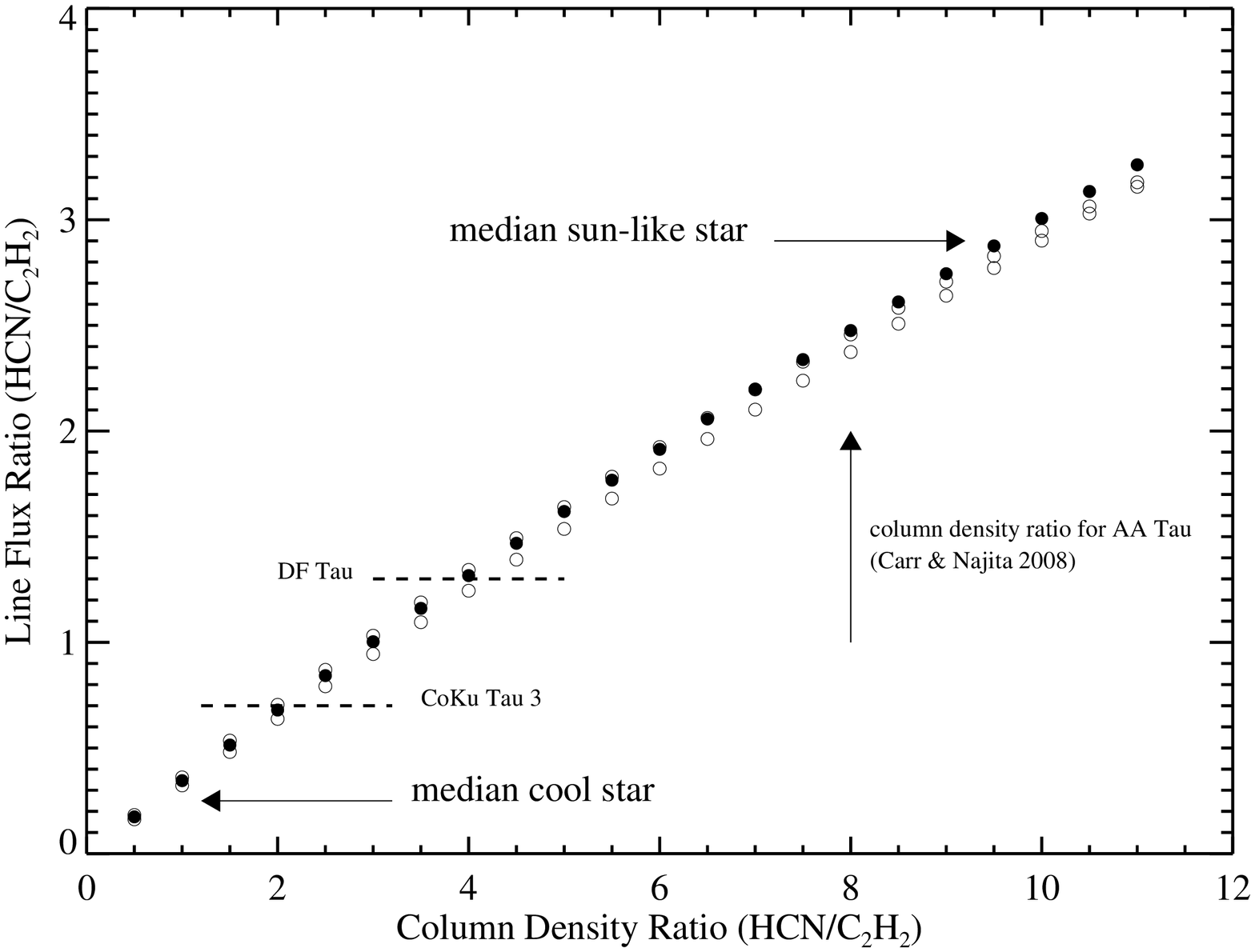}
\caption{Line flux ratio of HCN over \ctht{} versus column density ratio of the two molecules.
Circles are values predicted from sythetic spectra calculated following \citet{lahuis00}. The \ctht{} column density is kept fixed at 10$^{16}$\,cm$^{-2}$ while the HCN column density varies from 0.5-11$\times$10$^{16}$\,cm$^{-2}$, thus covering the \ctht{} and HCN column densities reported by \citet{carr08} for AA~Tau.
Filled circles are for a gas temperature of 650\,K while empty circles are for gas at 450 and 850\,K. 
Arrows indicate the line flux ratios measured from the median cool star and sun-like star spectra in Fig.~14. The dashed lines report the HCN over \ctht{} flux ratios for the only two sources in our sample where both molecular lines are firmly detected. The figure illustrates that the large difference in the HCN/\ctht{} flux ratio between the cool star and the sun-like star samples translates into very different column density ratios of the two molecules. We note that in the case of AA~Tau \citet{carr08} find an HCN/\ctht{} column density ratio of
 8, similar to what we require for the median spectrum of the sun-like stars.\label{fig:abun_fig}}
\end{figure}

\acknowledgments
We would like to thank S. Strom, J. Najita, and J. Carr for interesting discussions.
I. P. thanks T. Heran who contributed to the realization of Fig.~\ref{fig:Mdotvs10um} in the frame of the NASA Space Grant Program. K. L. was supported by grant AST-0544588 from the National Science Foundation.
This work is based on observations made with the Spitzer Space Telescope, which is operated by the Jet Propulsion Laboratory, California Institute of Technology. We are pleased to acknowledge support through the NASA/RSA contract number 1351891. M. R. M. acknowledges the support of the NASA Astrobiology Institute through LAPLACE.



{\it Facilities:} \facility{Spitzer Space Telescope}

\clearpage

\clearpage
\clearpage
\clearpage
\clearpage

\clearpage

\clearpage

\clearpage

\clearpage

\clearpage

\clearpage

\clearpage

\clearpage

\clearpage

\clearpage

\clearpage

\clearpage
\begin{deluxetable}{l cc ccc c}
\tabletypesize{\scriptsize}
\tablecaption{Sample of cool stars in the Cha~I star-forming region identified via infrared photometry. 
\label{T:photometry}}
\tablewidth{0pt}
\tablehead{
\colhead{Source} &\colhead{2MASS~J\tablenotemark{a}}  &\colhead{Sp. Type}  &
\colhead{$F_{6.7\mu m}$} & \colhead{$F_{14.3\mu m}$} & \colhead{$F_{8\mu m}$} &
\colhead{Other name}  \\
\colhead{} & \colhead{}  & \colhead{}  &
\colhead{(mJy)} & \colhead{(mJy)} & \colhead{(mJy)}  & 
\colhead{} 
}
\startdata
Hn~2$^\ast$ &11034764-7719563 & M5     &11.4$\pm$0.8& 3.7$\pm$1.0 & 9.1 & ChaI~425,ISO~32  \\
CHXR~15$^\ast$ &11054300-7726517 & M5.25  &13.2$\pm$1.1&  --        & 8.5 & ChaI~449,ISO~65  \\ 
ESO-H$\alpha$559&11062554-7633418 & M5.25     & -- & -- &  11.4 &  \\ 
ISO~79         &11063945-7736052 & M5.25  &8.0$\pm$2.0 & 9.9$\pm$1.5& 8.2 &    \\ 
CHSM~9484   &11071181-7625501 & M5.25  & -- & -- & 2.4  &ChaI~711 \\ 
Cha~H$\alpha$~1 &11071668-7735532 & M7.75  &6.7$\pm$1.8 &11.8$\pm$1.5& 8.1  &ISO~95   \\ 
Cha~H$\alpha$~9 &11071860-7732516 & M5.5   &11.4$\pm$0.9&9.2$\pm$1.3 & 9.4  & ISO~98  \\ 
Cha~H$\alpha$~2 &11074245-7733593 & M5.25  &24.5$\pm$1.4&23.4$\pm$1.2& 23.3  & ISO~111  \\
ISO~138     &11081850-7730408 & M6.5   &15.3$\pm$0.8&17.6$\pm$1.3&  1.8 & ChaI~410  \\ 
J11082570-7716396      &11082570-7716396 & M8     & --        & --        & 1.0 &    \\ 
ISO~147        &11082650-7715550 & M5.75  &4.3$\pm$1.1 & --        & 4.7 &   \\
Cha~H$\alpha$~6 &11083952-7734166 & M5.75  &12.7$\pm$1.5& -- & 11.8   & ISO~152    \\
J11084952-7638443    &11084952-7638443 & M8.75    & -- & -- & 0.8 &  \\ 
T37            &11085090-7625135 & M5.25  &7.5$\pm$1.5& -- & 11.1 & ISO~157 \\ 
ISO~165        &11085497-7632410 & M5.5   &12$\pm$1.8 & -- & 10.1 &   \\
ISO~217      &11095215-7639128 & M6.25  &10.3$\pm$2.5& --& 20.7  & ChaI~726     \\ 
ISO~252        &11104141-7720480 & M6     &5.5$\pm$0.8 &6.8$\pm$1.2& 8.1  \\ 	
Hn~13          &11105597-7645325 & M5.75  &31.4$\pm$1.8&27.5$\pm$1.9 & 44.4& ISO~259 \\ 
J11112249-7745427$^b$  &11112249-7745427 & M8.25 & -- & -- &	-- & \\ 
\enddata
\tablecomments{All spectral types are from \citet{luhman04,luhman08} and \citet{luhman05}. They are determined from optical spectra and have typical precisions of 0.25. Fluxes at 6.7 and 14.3\,\micron{} are from \citet{persi00}, while fluxes at 8\,\micron{} are from \citet{luhman08}. The 8\,\micron{} IRAC fluxes are mean values of fluxes from multiple campaigns.}
\tablenotetext{a}{The 2MASS source name includes the J2000 sexagesimal, equatorial position in the form: hhmmssss+ddmmsss \citep{cutri03}.}
\tablenotetext{\ast}{The IRS spectra are consistent with photospheric emission, see text.}
\tablenotetext{b}{This source has no 8\,\micron{} IRAC data but has excess emission at 5.8\,\micron}
\end{deluxetable}

\begin{deluxetable}{l  cc ccc}
\tabletypesize{\scriptsize}
\tablecaption{Log of the observations.\label{T:Observations}}
\tablewidth{0pt}
\tablehead{
\colhead{Source}&\colhead{Date}&\colhead{Peak--up}&  & \colhead{Exposure Times}&        \\
     		  &\colhead{mm/dd/year}    &\colhead{Mode}   &\colhead{SL1} &\colhead{SL2}  &\colhead{LL1}
}
\startdata
Hn~2		  & 03/12/2005    & PCRS  & 60s$\times$20    & -- &30s$\times$36	      \\
CHXR~15		  & 03/12/2005    & PCRS  & 60s$\times$20    & -- &30s$\times$36	      \\
ESO-H$\alpha$559 & 08/02/2006 & PCRS & 60s$\times$40 & 60s$\times$12 & -- \\
ISO~79              & 03/13/2007 & PCRS & 60s$\times$40 & 60s$\times$10 & -- \\
CHSM~9484	          & 08/02/2006 & PCRS & 60s$\times$40 & 60s$\times$12 & -- \\
Cha~H$\alpha$~1   & 03/12/2005    & PCRS  & 60s$\times$20& -- & --			    \\
Cha~H$\alpha$~9   & 03/12/2005    & PCRS  & 60s$\times$20& -- &30s$\times$36		     \\
Cha~H$\alpha$~2   & 03/12/2005    & PCRS  & 60s$\times$14   & -- & --			     \\
ISO~138    	  & 03/12/2005    & PCRS  & 60s$\times$14   & -- &--		      \\
J11082570-7716396        & 03/11/2007 & PCRS & 60s$\times$40 & 60s$\times$12 & -- \\
ISO~147            & 03/11/2007 & PCRS & 60s$\times$40 & 60s$\times$12 & -- \\
Cha~H$\alpha$~6   & 03/12/2005    & PCRS  & 60s$\times$20   & -- &30s$\times$36		     \\
J11084952-7638443     	& 03/11/2007 & IRS & 60s$\times$40 & 60s$\times$12 & -- \\
T37                   & 07/25/2006 & PCRS & 60s$\times$40 & 60s$\times$10 & -- \\
ISO~165           & 08/02/2006 & PCRS & 60s$\times$40 & 60s$\times$12 & -- \\
ISO~217		  & 03/12/2005    & IRS   & 60s$\times$20   & --&30s$\times$36	      \\
ISO~252          & 03/11/2007 & PCRS & 60s$\times$40 & 60s$\times$12 & -- \\
Hn~13               & 03/11/2007 & IRS & 60s$\times$40 & 60s$\times$12 & -- \\
J11112249-7745427     & 08/02/2006 & PCRS & 60s$\times$40 & 60s$\times$10 & -- \\
\enddata
\end{deluxetable}

\begin{deluxetable}{l cccccc}
\tabletypesize{\scriptsize}
\tablecaption{Statistics on the firm detections of HCN and \ctht{} from the sun-like and cool stars with disks. Because the sun-like stars were selected to have silicate emission features, we have excluded from the cool star sample the 3 featureless cool stars  (J11082570, 
J11084952, and J11112249). Also shown in the Table are the mean stellar temperatures, masses, radii, and accretion rates, used to compute the accretion luminosities. Values for individual stars are from \citet{apai05,luhman07,pas08,kh95,najita07,wb03,natta04,muzerolle05}.\label{T:det_stat}}
\tablewidth{0pt}
\tablehead{
\colhead{Sample}&\colhead{\ctht}&\colhead{HCN}&       
   \colhead{$<T_\star>$}    &\colhead{$<M_\star>$}   &\colhead{$<R_\star>$} &\colhead{$<\dot{M}>$} \\
\colhead{} & \colhead{detections} &  \colhead{detections} &
   \colhead{[K]}    &\colhead{[M$_\sun$]}   &\colhead{[R$_\sun$]} &\colhead{M$_\sun$/yr} 
}
\startdata
Sun-like stars & 4/44 & 13/44 & 3,900 & 0.8 & 2.1 & 1.3$\times 10^{-8}$ \\
Cool stars  & 5/14 & 0/14 & 3,000 & 0.1 & 0.7 & $<$3$\times 10^{-11}$ \\
\enddata
\end{deluxetable}

\end{document}